\documentclass[twocolumn,iop,aps,superscriptaddress,showpacs,floatfix]{revtex4-1}

\usepackage{mathptmx}
\usepackage{subfigure}
\usepackage{dcolumn}
\usepackage{amsmath,amssymb}
\usepackage{bm}
\usepackage{color}
\usepackage{latexsym}
\usepackage{epstopdf}
\usepackage{color}
\usepackage[english]{babel}
\usepackage{latexsym}

\usepackage{psfrag,graphicx} 
\usepackage{epsf} 
\usepackage{subfigure} 
\usepackage{amsfonts}
\usepackage{bm}
\usepackage{natbib}
\usepackage{epstopdf}
\usepackage{appendix}
\usepackage{changes}
\usepackage{bbm}

\definecolor{mygrey}{gray}{0.35}
\definecolor{myblue}{rgb}{0.2,0.2,0.8}
\definecolor{myzard}{cmyk}{0,0,0.05,0}
\definecolor{mywhite}{rgb}{1,1,1}
\definecolor{myred}{rgb}{1,0.,0.3}

\usepackage[colorlinks=true,citecolor=myblue,linkcolor=myred]{hyperref}

\def\be{\begin{equation}}
\def\ee{\end{equation}}
\def\ba{\begin{align}}
\def\enda{\end{align}}
\def\bi{\begin{itemize}}
\def\ei{\end{itemize}}

 \def\ee{\mathord{\rm e}}

 \def\ee{\mathord{\rm e}}

\renewcommand{\ee}{{\rm e}}

\def\beq{\begin{equation}}
\def\beq{\begin{equation}}
\def\eeq{\end{equation}}

 \newcommand{\ket}[1]{|#1\rangle}

\begin{document}

\title[Short Title]{Quantum spectroscopy of single spins assisted by a
classical clock}
\author{Tuvia Gefen}
\affiliation{Racah Institute of Physics, The Hebrew University of Jerusalem, Jerusalem 
91904, Givat Ram, Israel}
\author{Maxim Khodas}
\affiliation{Racah Institute of Physics, The Hebrew University of Jerusalem, Jerusalem 
91904, Givat Ram, Israel}
\author{Liam P. McGuinness}
\affiliation{Institute for Quantum Optics, Ulm University, Albert-Einstein-Allee 11, Ulm 89081, Germany}
\author{Fedor Jelezko}
\affiliation{Institute for Quantum Optics, Ulm University, Albert-Einstein-Allee 11, Ulm 89081, Germany}
\author{Alex Retzker}
\affiliation{Racah Institute of Physics, The Hebrew University of Jerusalem, Jerusalem 
91904, Givat Ram, Israel}
\date{\today}


\begin{abstract}
Quantum spectroscopy with single two level systems has considerably improved our ability to detect weak signals.
Recently it was realized that for classical signals, precision and resolution of quantum spectroscopy is limited mainly by coherence of the signal and stability of the clock used to measure time.
The coherence time of the quantum probe, which can be significantly shorter, is not a major limiting factor in resolution measurements.
Here, we address a similar question for spectroscopy of quantum signals, for example a quantum sensor is used to detect a single nuclear spin.  
We present and analyze a novel correlation spectroscopy technique with performance that is limited by the coherence time of the target spins and the stability of the clock. 
\end{abstract}
\maketitle

\subsection {Introduction}
Quantum metrology and quantum sensing \cite{degen2016,giovannetti2011advances} are extremely promising research directions which use quantum mechanics to reach the ultimate limits of measurements accuracy. One of the major goals of this field is the measurement of magnetic fields. State-of-the-art magnetometry often relies on dynamical decoupling where fast pulses or continuous fields drive a quantum mechanical system\cite{hahn1950spin,viola1998dynamical,biercuk2009optimized,hall2010ultrasensitive,kotler2011single,taylor2008high,balasubramanian2008nanoscale}. The role of these fields is to decouple the system from the environment, and thus to enhance the coherence time ($T_2$), while at the same time retaining the ability to sense a signal that is on resonance with the pulse rate. 

Characterizing the time dependence of signals, either quantum or classical \cite{de2016estimation,gefen2017control,pang2016quantum,yang2016quantum} has become one of the central goals of quantum sensing in the last few years. The interest in this research stems from the ability to sense frequencies with increased resolution.
In particular, in recent years correlation spectroscopy has been extensively used in the field of  NV centers in diamond\cite{laraoui2013high,zaiser2016enhancing,staudacher2015probing,ajoy2015atomic,rosskopf2016quantum,schmitt2017submillihertz,laraoui2011diamond,pfender2016nonvolatile}. Notably, one of these schemes \cite{schmitt2017submillihertz,boss2017quantum,bucher2017high} utilized a classical clock rather than a quantum memory. We now present a study based on this work, which exploits the ability to manipulate classical data in order to construct a protocol that targets quantum signals with improved sensitivity.

We address the problem of a quantum probe that interacts with a target system aiming to estimate its energy gaps. When such energy gaps arise from local couplings in the quantum system (eg chemical shifts), their estimation provides detailed information about the physical structure of the system.
We address two issues: 
1) Given that only one nearby quantum system has a frequency in a certain range, we seek to estimate this frequency. We refer to this task as precision. 2) Knowing that there are a few nearby quantum systems with similar frequencies we aim to resolve their energy gaps. We term this task, resolution.
In this manuscript, we present and analyze a scheme in which the precision and resolution are limited by the coherence time of the target system and not the coherence time of the probe. 
Although our protocol is general, we focus on the case of nitrogen-vacancy (NV) centers in diamond.
The NV center is coupled to nearby $^{13}C$ nuclear spins in diamond (fig .\ref{scheme} (a)), and our goal is to estimate the Larmor frequency of some of these nuclei.
The main limitation is that these nuclear spins cannot be measured or polarized directly, and the only way to apply these operations is through the NV center {\footnote{We remark that nuclear spins can be manipulated by applying rf pulses on them. This alone does not provide the ability to measure or polarize them. }}.

\subsection{Simplified Model}
The main idea of the protocol is best explained by the following simplified model, which is very similar to the effective Hamiltonian in the NV scenario:
\begin{equation}
H_{S}=\omega_{l} I_Z +\tilde{g} \sigma_z I_X,
\end{equation}
where $I_K$ is the $K$'th component of the nuclear spin and $\sigma_z$ is the $z$ component of the electron spin. We wish to estimate $\omega_{l},$ the Larmor frequency of the nuclear spin.
We first address the relevant case in which: $\tilde{g} \ll \omega_{l},$ and the nuclear spin is completely unpolarized, i.e. its initial state is: $\rho_{n}=\frac{1}{2} \mathbbm{1}.$ 
Therefore in order to obtain information about $\omega_{l},$ the nuclear spin must first be polarized. Then it should be allowed to rotate due to $\omega_{l} I_{Z},$ and eventually a measurement should be applied.
Note that the measurement and polarization are both obtained through the interaction term: $\tilde{g} \sigma_z I_x,$ while the rotation is due to $\omega_{l} I_{Z}.$ The protocol is therefore divided into three parts: Two detection periods, in which $\tilde{g} \sigma_z I_x$ should be dominant, one for initialization and one for measurement, and a rotation period in which $\omega_l I_z$ should be dominant.
In order to make $\tilde{g} \sigma_{z} I_{X}$ dominant in the detection periods, an external drive is applied to the electron spin. The standard choice would be a train of $\pi$-pulses with a spacing as close as possible to $\omega_{l},$ thus the Hamiltonian in the detection periods reads:
\begin{equation}
H_{S2}=\omega_{l} I_Z + \tilde{g} \sigma_z I_X+\Omega \left( t \right) \sigma_{x}.
\end{equation} 
$\Omega \left( t \right)$ consists of a set of sharp $\pi$-pulses, approximated as delta functions, with a spacing of $\tau=\frac{\pi}{\omega_e},$ where $\omega_{e}$ is our estimation of $\omega_{l}.$
To show that this effectively reduces $\omega_{l},$ let us consider the interaction picture with respect to these pulses and $\omega_{e} I_{z},$ where we obtain:
 \begin{equation}
H_{S2I}=\delta I_{Z}+\tilde{g} h\left(t\right)\sigma_{z}\left( I_{X}\cos\left(2 \omega_{e}t\right)+I_{Y}\sin\left(2 \omega_{e}t\right) \right), 
\end{equation} 
 where $h \left( t \right)$ is the square-wave:
 \begin{equation}
 	h(t)=\begin{cases} -1 &  2\omega_e t     \,\, \left(  \text{mod}  \,\,  2 \pi   \right)  \in (\frac{\pi}{2},\frac{3 \pi}{2})\\
					1 &    \text{else} .\\	
	\end{cases}
\label{square-wave}	
\end{equation} 
Observe that:
\begin{equation}
h\left(t\right)=\frac{4}{\pi}\underset{\text{odd n}}{\sum}\frac{1}{n}\cos\left(2 n \omega_{e}t\right).
\label{Fourier square-wave}
\end{equation}
 Since $\tilde{g} \ll \omega_{l},$ and assuming that $\delta:=\omega_{l}-\omega_{e} \ll \omega_{l},$ we can neglect all the fast rotating terms (those with rotation frequency that goes as $\omega_{l}$), using eq. \ref{Fourier square-wave} it can be seen that we are left with:
\begin{equation}
H_{S2I}\approx\frac{2}{\pi} \tilde{g} \sigma_{z} I_{X} +\delta I_{Z}.
\end{equation}
Given that $\delta \ll \tilde{g},$ the interaction is more dominant than the rotation, in fact in the limit of $\delta T_{2} \ll 1,$ we can approximate:
\begin{equation}
H_{S2I}\approx\frac{2}{\pi} \tilde{g} \sigma_{z}I_{X}.
\end{equation}
This is exactly the Hamiltonian required to measure of the nuclear spin along the X-axis\cite{jacobs2006straightforward,wiseman2009quantum}: The NV center is initialized to $|\uparrow_{y}\rangle_{e}$ and measured in the $\sigma_{X}$-basis.
 It should be noted that if the coherence time of the probe is short (compared to the coupling), only a weak measuremeent can be performed. This scheme of measuring the nuclear spin weakly through the electron spin was in fact already analyzed and implemented in an experiment \cite{liu2017single}. 
Between these two detection periods, the system evolves freely according to $H_{S}=\omega_{l} I_Z +\tilde{g} \sigma_z I_X \approx \omega_{l} I_{Z},$ therefore the nuclear spin is rotated with frequency of $\omega_{l}.$
This is basically a standard Ramsey experiment in which only weak initialization and measurement are possible. It can be then seen that the probability for positive (negative) correlation is given by: $P_{+}=\frac{1}{2}+\frac{1}{2}\sin\left(2g\tau\right)^{2}\cos\left(2\omega_{l}t\right)$ $\left(P_{-}=\frac{1}{2}-\frac{1}{2}\sin\left(2 g\tau\right)^{2}\cos\left(2\omega_{l}t\right)\right)$, where $g=\frac{2 \tilde{g}}{\pi}$ and $\tau$ is the duration of the detection period, from which the Larmor frequency can be estimated. 
 
\subsection {Description of the scheme for NV center}
\subsubsection{Single nucleus}
\begin{figure}[h]
\begin{center}
\includegraphics[width=0.4\textwidth]{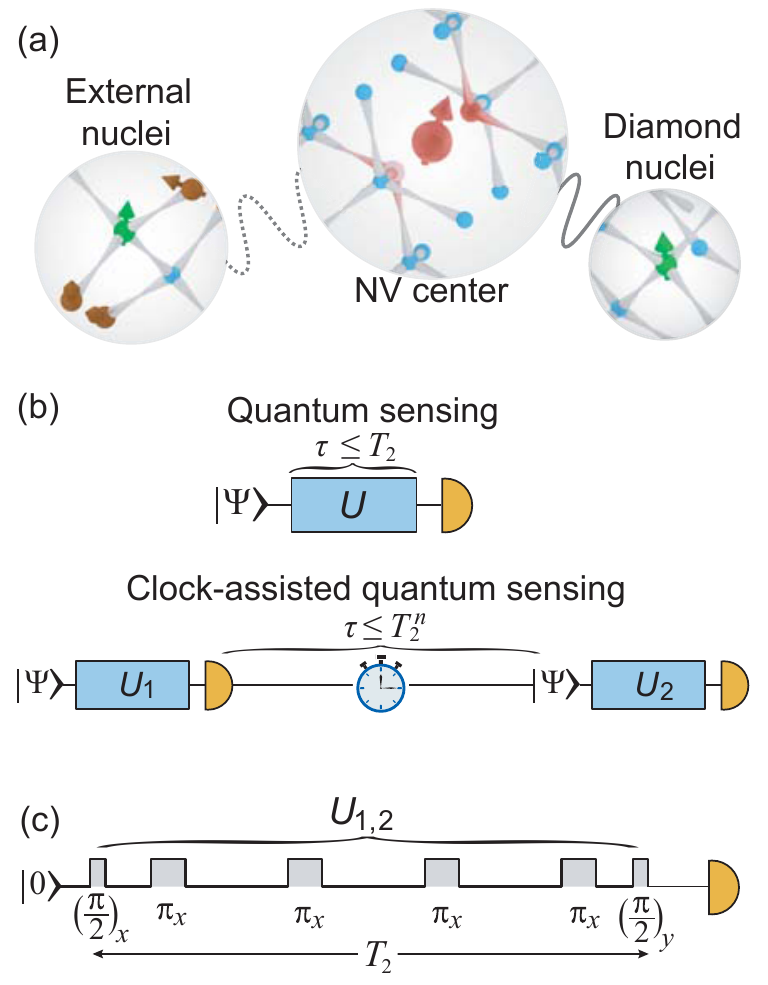}
\end{center}
\caption{(a) {\it The central problem.} A quantum sensor which is based on an NV center aims to resolve two quantum systems with very close Larmor frequencies in the presence of classical noise, either inside the bulk diamond or an external spin. (b) Conventional quantum sensing uses pulse sequences with different frequencies to track the desired Larmor frequencies. $\ket \Psi$ denotes the initial state and $U$ stands for the time evolution of the NV. The resolution of this method is limited by the $T_2$ of the NV.  The method we propose is based on the observation that correlations exist already in measurements outcomes and thus the quantum state does not need to be stored in a memory qubit. This method is thus limited by the coherence time of the signal, namely the nuclear spin, or the clock stability. (c) All methods use a dynamical decoupling sequence, denoted above as $U_{1,2},$ which reduces the effect of noise and probes a frequency which is close to the frequency of the pulse sequence.}
\label{scheme}
\end{figure}

We consider the case of a single NV center coupled to nearby $^{13}C$ nuclear spins in diamond (fig .\ref{scheme} (a)), and the objective is to estimate the Larmor frequency of one of these nuclei.
Heuristically, thanks to the coupling between the NV and the $^{13}C,$ measurement of the NV electronic spin induces a measurement of the $^{13}C$ nuclear spin\cite{jacobs2006straightforward,wiseman2009quantum}. 
We investigate the scenario where the nuclear spin is far from the NV center, such that the coupling is short compared to the coherence time of the NV.
The protocol then consists of two weak measurements: one at the beginning and one at the end. The first measurement creates a small coherence (polarization) in the $^{13}C,$ and after a period of $\tau$ a second measurement provides information about the rotation frequency.   
 As the quantum sensor, the NV center, plays no role during the time $\tau,$ this scheme is limited only by the coherence time of the nuclear spin and stability of the clock.

In the following we explain the protocol in more detail. The Hamiltonian of the system is
\begin{equation}
H = D S_z^2 - \gamma_e B_z S_z  - \sum_j \gamma_{n,j} B_z I_{z,j} +\frac{1}{2} \sum_{jk} A_{zk} S_z I_{k,j},
\label{H1}
\end{equation}
where $S_z$ is  the $z$ component of the spin one of the NV center, $I_{k,i}$ is the $k$-component of the spin of the $j$-th nucleus, $D$ is the NV zero field splitting, $\gamma_e, \gamma_{n,i}$ are the gyromagnetic ratios of the NV and the $j$-th nuclear spin respectively,
 and $A_{zk}$ is the hyperfine coupling between the NV and the nuclei.
 In Eq.~\eqref{H1} the index $j$ runs over the nuclei and $k$ over the directions. 
Due to the external magnetic field and the zero field splitting, $S_z=0$ and $S_z=-1$ states can be coupled, while leaving $S_z=1$ untouched. Therefore the $S_z=1$ state becomes irrelevant and the above two states form a qubit with energy levels described by: $\omega_{0} S_Z=\omega_{0} \frac{1}{2} \left( \sigma_{z} - \mathbbm{1}  \right),$ where $\omega_{0}=D-\gamma_{e} B_{z}.$ 
Hence, we describe the above two coupled states as the eigenstates of the spin half operator $S_{Z}=\frac{1}{2} \left( \sigma_{z}-\mathbbm{1} \right)$.

In Eq. \eqref{H1} we have made the secular approximation and neglected all the terms containing $S_{x,y}$. 
Furthermore, we have omitted the noise terms due to magnetic field fluctuations operating on the NV center and the nuclei. 
These terms limit the coherence times, $T_2$ and $T_{2}^{n}$ of the NV(after dynamical decoupling) and the nuclei respectively. 

The Hamiltonian, with an appropriate choice of axes and assuming a single nuclear spin, reads:
\begin{equation}
H=\omega_{0} S_{Z}+ \omega_l I_z + \tilde{g} S_z I_x+A S_{z} I_{z}.
\end{equation}
The protocol is again divided into three parts: two detection periods, one for initialization and one for final measurement, in which $\tilde{g} S_z I_x$ should be dominant, and rotation period in which $\omega_l I_z$ should be dominant.

Let us focus first on the two detection periods: Typically $\omega_{l} \gg \tilde{g},$ therefore in order to make $\tilde{g} S_z I_x$ dominant, we must apply an additional drive: A sequence of $\pi-$pulses ($XY8$ sequence for example).
Note that due to the  term of: $A S_z I_z,$ the Larmor frequency is shifted, and the spacing between the $\pi$-pulses should correspond to this shifted frequency.
The Hamiltonian in these two detection periods is therefore:
\begin{equation}
H=\omega_{0} S_{Z}+\Omega(t) \sigma_{X}\cos(\omega_{0}t) + \omega_l I_z + \tilde{g} S_z I_x+A S_z I_z.
\end{equation}
The XY8 sequence $\Omega(t)$ consists of a set of sharp $\pi$  pulses, approximated as delta functions, with a spacing of $\tau_p=\frac{\pi}{\omega_{e}}$ from one another, where $\omega_{e}$ is our estimation of $\omega_l-\frac{A}{2}$. We claim that, like in the simplified model, this should effectively reduce $\omega_{l}.$        
Moving to the interaction picture with respect to $\omega_{0} S_z,$ and assuming $\Omega(t)\ll\omega_{0},$ we get in the rotating wave approximation:
\begin{equation}
H_{I}=\Omega(t) \sigma_{X} + \omega_l I_z + \tilde{g} S_z I_x+A S_z I_z.
\end{equation}

Moving now to the interaction picture with respect to the pulses we get:
\begin{eqnarray}
\begin{split}
&H_{II}=\omega_{l}I_{Z}+\frac{\tilde{g}}{2}\left(h\left(t\right)\cdot\sigma_{z}-\mathbbm{1}\right)I_{X}+\frac{A}{2}\left(h\left(t\right)\cdot\sigma_{z}-\mathbbm{1}\right)I_{Z}\\
&=\left(\omega_{l}-\frac{A}{2}\right)I_{Z}+\frac{\tilde{g}}{2}\left(h\left(t\right)\cdot\sigma_{z}-\mathbbm{1}\right)I_{X}+\frac{A}{2}h\left(t\right)\sigma_{Z}I_{Z},
\end{split}
\end{eqnarray}  
where $h(t)$ is the same square wave function as in eq. \ref{square-wave}.
We assume that $\omega_{e}$ is very close to $\omega'=\omega_l-\frac{A}{2},$ i.e. $\delta=\omega'-\omega_{e}\ll \omega'.$

\begin{figure*}
\begin{center}
\includegraphics[width=0.9\textwidth]{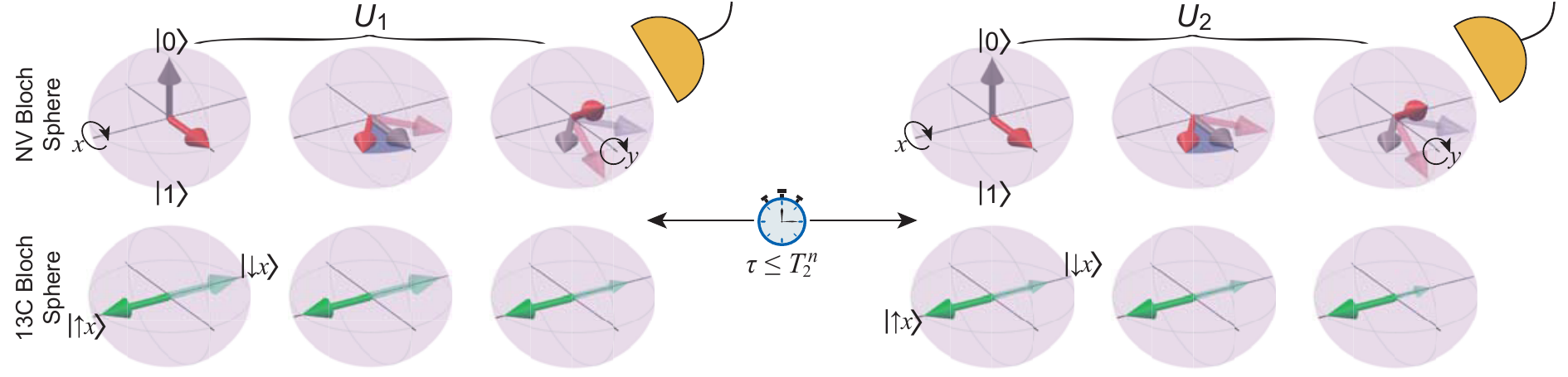}
\end{center}
\caption{{\it The explanation of the operation principle of the protocol.}
Description of the complete measurements cycle.
Upon initialization the NV spin (red) is polarized along $|\uparrow_y \rangle_{e} $ and the $^{13}C$ nucleus is in a state of identity density matrix.
 In the first detection period ($U_{1}$), half of the NV population precesses clockwise and the other half counterclockwise around the $z$ axis.
Measuring in the $\sigma_{x}$ basis of the NV induces a finite polarization of $^{13}C$ along $\hat{x}$.
During the rotation period, after time of $ \tau = n \pi \omega^{-1}$ the polarization of $^{13}C$ acquired at the first measurements causes the electron spin to precess more clockwise than counterclockwise. 
This imbalance is revealed by a second measurement in $\sigma_{x}$ basis of the NV.
In contrast, if the time interval between the two measurements is $\tau = (n/2 +1/4 )\pi \omega^{-1}$
the precession of NVs is determined by nuclei polarization along $\hat{y}$.
Since equal amount of $^{13}C$ is polarized along $\vert \uparrow_{y} \rangle_{e}$ and $\vert \downarrow_{y} \rangle_{e},$ 
there is no excess polarization of the NV in this case.}
\label{scheme_Ramsey}
\end{figure*}
\noindent

Moving to the interaction picture with respect to $\omega_{e} I_z$ we get:
\begin{eqnarray}
\begin{split}
&H_{III}= \delta I_{z}+\frac{\tilde{g}}{2}\left(h\left(t\right)\cdot\sigma_{z}-\mathbbm{1}\right)\left(I_{X}\cos\left(2 \omega_{e}t\right)+I_{Y}\sin\left(2 \omega_{e}t\right)\right)\\
&+\frac{A}{2}h\left(t\right)\sigma_{Z}I_{Z},
\end{split} 
\label{main}
\end{eqnarray}
since $\tilde{g},A \ll \omega_{e},$ we can neglect all the fast oscillating terms (those that oscillate like $\omega',\omega_{e}$) and stay with: 
 \begin{equation}
 H_{\text{eff}}= \delta I_{Z} + g \sigma_z I_{X} \approx g \sigma_z I_{X} ,
 \label{effective_detection}
 \end{equation}
where $g = \tilde{g}/\pi$.

We therefore obtained the required Hamiltonian for the two detection periods, in between these two periods we would like the nuclear spin to rotate. To that end we should change our control,
such that the rotation would be dominant. This can be achieved by applying $\pi$-pulses with a frequency $\omega_{c},$ such that: $\omega_{c},\:\omega'-\omega_{c}\gg\widetilde{g},A,$ in that case the effective Hamiltonian is given by:
\begin{equation}
H_{\text{eff}}= \omega I_{Z}+g\sigma_{z}I_{X}\approx \omega I_{Z},
\label{effective_rotation}
\end{equation}
where $\omega=\omega'-\omega_c.$
We can now conclude, the protocol reads as follows:

\begin{center}
\fbox{ \begin{minipage} [t] {0.42 \textwidth}
{\bf{The Protocol:}}
\\
 First detection period: Initialize the electron spin to $|\uparrow_{Y}\rangle_{e}.$ Apply a sequence of $\pi$-pulses with a spacing of $\pi/\omega_{e},$ where $\omega_{e}$ is the estimation of $\omega'.$ 
The obtained effective Hamiltonian is the one in eq.\ref{effective_detection}. After a period of $\tau_{m},$ measure the electron spin in $\sigma_{x}$ basis.
\\
\\
Rotation period: Change the frequency of the pulses to $\omega_{c},$ the obtained effective Hamiltonian is now the one in eq. \ref{effective_rotation}. No importance as to what is the initial state of the electron spin. The length of this period, denoted as $\tau$ is limited only by the coherence time of the nuclear spin.
\\
\\
Second detection period: The same as the first detection period.
\end{minipage}
}
\end{center}

Given these approximations, it can be seen that the  probability for a positive (negative) correlation between the first measurement outcome and the second measurement outcome is given by:   
\begin{align}
p_{\pm}=\frac{1}{2}\left[1\pm\sin^{2}(2 \phi)\cos(2\omega\tau)\right],
\label{corr1}
\end{align}
where $p_{+}$ ($p_{-}$) denotes positive (negative) correlation,
$\tau$ is the length of the rotation period and the dimensionless coupling $\phi$ stands for $\phi = g \tau_m.$
The latter can be understood as follows: if $\omega \tau=\pi$ the nuclear spin gets back to its original state and there is a high correlation, while the contrast is set by the coupling strength.
A more detailed explanation of the probability is the following: a measurement of the NV at the first detection period projects the  NV to the state $\vert \uparrow_x \rangle_{e}$ and thus the nucleus collapses to: 
\begin{eqnarray}\label{imb}
\rho_n &=& \frac{1}{2}\left[ 1- \sin(2 \phi) \right] \vert \downarrow_x \rangle_{n}  \, _{n}\langle \downarrow_x \vert 
\notag \\
& +& \frac{1}{2}\left[ 1+ \sin(2 \phi) \right] \vert \uparrow_x \rangle_{n} \, _{n} \langle \uparrow_x \vert,
\end{eqnarray}
i.e., an imbalance between the $\vert \uparrow_x \rangle_{n}$ and $\vert \downarrow_x \rangle_{n}$ states, see Fig.~\ref{scheme_Ramsey}. 
The second measurement builds on this imbalance to achieve correlation with the first measurement.
The outcome of the second measurement depends on the state of nuclear polarization.
Given that the nucleus is in the $\vert \uparrow_x \rangle_{n}$ state,  the NV is measured in the $\vert \uparrow_x \rangle_{e}$ state with a probability of $\frac{1}{2} \left(1+ \sin \left(2 \phi \right) \cos (2 \omega  \tau )\right)$.
For the opposite nucleus polarization the NV is measured in the  $\vert \uparrow_x \rangle_{e}$ state with probability $\frac{1}{2} \left(1-\sin \left( 2 \phi \right) \cos (2 \omega  \tau )\right)$.
This leads to the correlation function presented in eq. \eqref{corr1}.
The periodicity in $\omega \tau$ can be simply understood.
Note that the precession direction of the electron spin is determined by the nuclear spin state.
For $\omega \tau =n \pi $ the nuclear spin polarization returns to its state after the first measurement, and thus a positive correlation is more likely.
For $\omega \tau =(2n+1) \frac {\pi}{2},$ however, the excess polarization generated by the first measurement is now at the opposite direction, which means that negative correlation is more likely. This is illustrated in fig. \ref{scheme_Ramsey}.

Given the probability in eq. \ref{corr1}, one obtains an uncertainty:
\begin{equation}
\Delta\omega=\frac{\sqrt{1-\sin^{4}\left( 2 \phi \right)\cos^{2}\left(2 \omega\tau\right)}}{2 \sin^{2}\left(2 \phi \right)\tau|\sin\left(2 \omega\tau\right)|}.
\label{precision1}
\end{equation} 
This result can be simply understood: In the relevant regime of weak coupling, $\phi \ll 1,$ we get for $\omega\tau=\left(2n+1\right)\frac{\pi}{4}$,
\begin{equation}
\Delta\omega=\frac{1}{8 \phi^{2}\tau}.
\label{precision2}       
\end{equation}      
Therefore a $\frac{1}{\tau}$ scaling is acieved, just like in a standard Ramsey experiment\cite{itano1993quantum,budker2007optical}, however,
since the measurement is weak, there is an additional small dimensionless pre-factor, $\phi^{2},$ which reduces the precision. 
The standard Ramsey scaling is retrieved for larger $\phi,$ i.e.,   $\phi=\left(2n+1\right) \frac{\pi}{4}$, where we get
$\Delta \omega=1/\left( 2\tau \right).$
Note that $\tau,$ the length of the rotation period, is limited only by the coherence time of the nuclear spin ($T_{2}^{N}$). 
The coherence time of the electron spin ($T_{2}$), on the other hand, appears only in $\phi.$ 
Therefore $T_{2}$ does not play a crucial role in the strong coupling regime and it is noteworthy that prolonging $T_2$ beyond $\frac{\pi}{4 g}$ does not improve the sensitivity which is analogous to the fact that in a regular Ramsey measurement stronger laser power does not improve precision.
 We can thus conclude by saying that in the strong coupling regime we achieve a standard Ramsey experiment with a length of $T_{2}^{n}$ (coherence time of nuclear spin), while in the weak coupling regime, using only two weak measurements we have: 
 \begin{equation}
 \Delta \omega= \frac{1}{8 \phi^{2} T_{2}^{N} \sqrt{N}},  
\end{equation}
where $N$ is the number of experiments performed.

\subsubsection {Several nuclei} 
In order to study resolution we can consider two nuclei. 
In this case we have a single NV center which interacts with two nuclei having two different detunings from the NV drive
$\omega_1 = \omega_{l}^{1}-\omega_{c},$  $\omega_2 = \omega_{l}^{2}-\omega_{c}$.
The probability for a positive (negative) correlation reads:
\begin{eqnarray}
\begin{split}
& p_{\pm}=\frac{1}{2}\pm\frac{1}{2}\cos(2\omega_{1}\tau)\sin^{2}\left(2\phi_{1}\right)\cos^{2}\left(2\phi_{2}\right) \\
&\pm\frac{1}{2}\cos(2\omega_{2}\tau)\cos^{2}\left(2\phi_{1}\right)\sin^{2}\left(2\phi_{2}\right),
\end{split}
\end{eqnarray}
where $\phi_1,\phi_2$ are defined as $g_1 \tau_m,g_2\tau_m$ respectively.
The $\cos^{2}(2 \phi_i)$ terms are due to interference of the contributions of the two nuclei.
We observe that in the weak coupling limit ($\phi_1,\phi_2\ll1$) the signals are independent,
\begin{equation}\label{p_res}
p_{\pm}=\frac{1}{2}\pm2\phi_{1}^{2}\cos(2\omega_{1}\tau)\pm2\phi_{2}^{2}\cos(2\omega_{2}\tau).
\end{equation}
This expression implies that the resolution goes as $1/\tau,$ which is limited by the coherence time of the nuclear spin ($T_{2}^{n}$). 
This can be understood in the following way: We can repeat the correlation experiment with different values of $\tau$, where the longest corresponds to $T_{2}^{n}$. A Fourier transform
of the outcomes will yield peaks at $\omega_{1},\omega_{2},$ each one with a linewidth of $1/T_{2}^{n},$ and thus a resolution of $1/T_{2}^{n}.$ 
The coherence time of the NV appears only in the parameters $\phi_{1,2},$ which determine the signal to noise ratio (SNR): In Fourier analysis the SNR of this experiment is given by $\phi_{i}^{2} \sqrt{N},$ where $N$ is the number of measurements.
Namely, peaks will appear only if $\phi_{i}^{2} \sqrt{N}>1.$ By taking a large enough $N,$ we can guarantee this condition holds. Therefore the NV coherence time does not serve as a fundamental limitation on resolution.   

Eq. \eqref{p_res} is readily generalized to the case of $n$ nuclei,
\begin{equation}\label{SxSx}
p_{\pm}=\frac{1}{2}\pm2\sum_{k=1}^{n}\phi_{k}^{2}\cos(2\omega_{k}\tau).
\end{equation}
Correspondingly, our estimates of the resolution apply to the general case.

There are a few important points that should be noted about the scheme. First, 
The NV may inflict noise on the nuclei by two different mechanisms. 
The $T_1$ process of the NV will induce a random field on the nuclei. 
This could be countered by driving the NV with a microwave as described previously in \cite{cohen2017protecting,wolfowicz201629si} or the defect ionization as in\cite{dreher2012nuclear,pla2013high,saeedi2013room,maurer2012room}.
Another option is to weakly irradiate the NV and thus constantly initialize the NV in the $m_s = 0$ state\cite{chen2017dissipatively}.
The nuclei may also be subjected to noise during the measurement and initialization process. This effect could be reduced by driving both the excited and the ground state in a correlated manner by extending the protocol described in \cite{cohen2017protecting}.

 An additional point concerns the first measurement in the scheme. As already described, the only objective of this measurement is to polarize the nuclear spin in a given direction at the $X-Y$ plane. 
In some cases it is possible to polarize the nucleus coherently at the beginning \cite{london2013detecting}, which has several advantages over measurement based state preparation.
First, it requires much less scattering of photons which will result in less noise on the nuclei. Second, since no correlation is measured, all the classical noise is averaged to zero.
It also has the advantage that the initial measurement process does not subject the nuclei to noise.

\subsection{Enhancing precision by repetitive measurements}  
\begin{figure}
\begin {center}
\includegraphics[width=6.2cm]{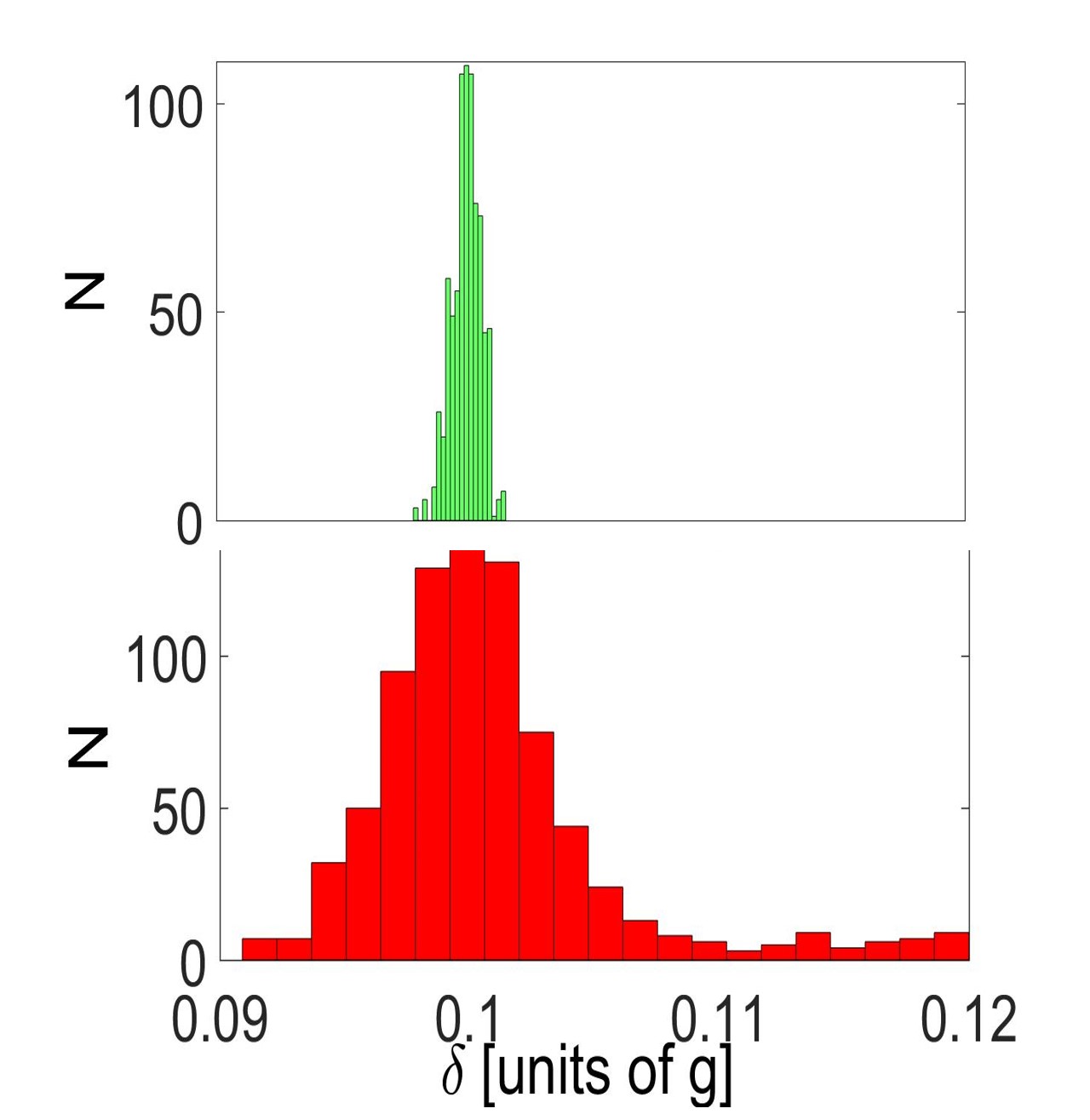}
\caption{ 
Different precision levels achieved with a different number of measurements. The figure presents histograms of the estimator of $\delta,$ achieved from simulating the scheme with parameters of $T=30 \, [g^{-1}], \, \phi=0.2, \, \delta=0.1 \, [g] .$  
The green (upper) histogram obtained with 15 measurements at each detection period ($n_{1}=15, \, n_{2}=15$), and the red (lower) histogram obtained with 2 measurements at each detection period ($n_{1}=2, \, n_{2}=2$).
Maximum Likelihood was used to estimate the frequency. The variance of these histograms, which quantifies the precision, is indeed $1/I$, where $I$ is the Fisher information.}
\label{histograms_multiple}
\end {center}
\end{figure}

So far we have addressed the case of a single measurement at the beginning and a single measurement at the end. For small $\phi,$ these are weak measurements which provide less information than a standard strong measurement. Indeed, as $\phi$ gets smaller, the precision drops as $\sin \left( 2 \phi  \right)^{2}$ (Eq. \ref{precision1}).
This should imply that for small $\phi$ it could be preferable to make further measurements, i.e. a sequence of measurements in every detection period.
In that case the protocol reads: $\text{(detection)}^{n_{1}}$-evolution-$\text{(detection)}^{n_{2}},$ i.e. $n_{1}$ measurements in the first detection period, followed by free evolution of the nuclear spin and $n_{2}$ measurements in the second detection period.
 Multiple measurements indeed increase precision,  as can be observed in fig. \ref{histograms_multiple}. 
This observation raises questions about the optimal number of measurements and the limits of the precision.  

The precision analysis of this case is a bit more involved, and requires the use of Cram{\'e}r-Rao bound. According to the Cram{\'e}r-Rao bound, the variance of any unbiased estimator of $\omega,$ is lower bounded by the inverse of the Fisher information: $\text{\text{Var}}\left(\hat{\omega}\right)\geq I^{-1},$ where $I$ denotes the Fisher information (FI).
We thus use the FI to quantify the precision, which makes the analysis as general as possible. The FI of $\omega$ is defined as: $I=\langle\left(\partial_{\omega}\ln\left(p\right)\right)^{2}\rangle=\underset{p}{\sum}\frac{\left(\partial_{\omega}p\right)^{2}}{p},$ where $p$ denotes the probability for a certain outcome, and the sum is over all the possible outcomes.
Thus in our scheme, one should sum over all the possible results of the NV measurements.

\begin{figure}
\begin {center}
\subfigure[]{\includegraphics[width=7.2cm,height=2.7cm]{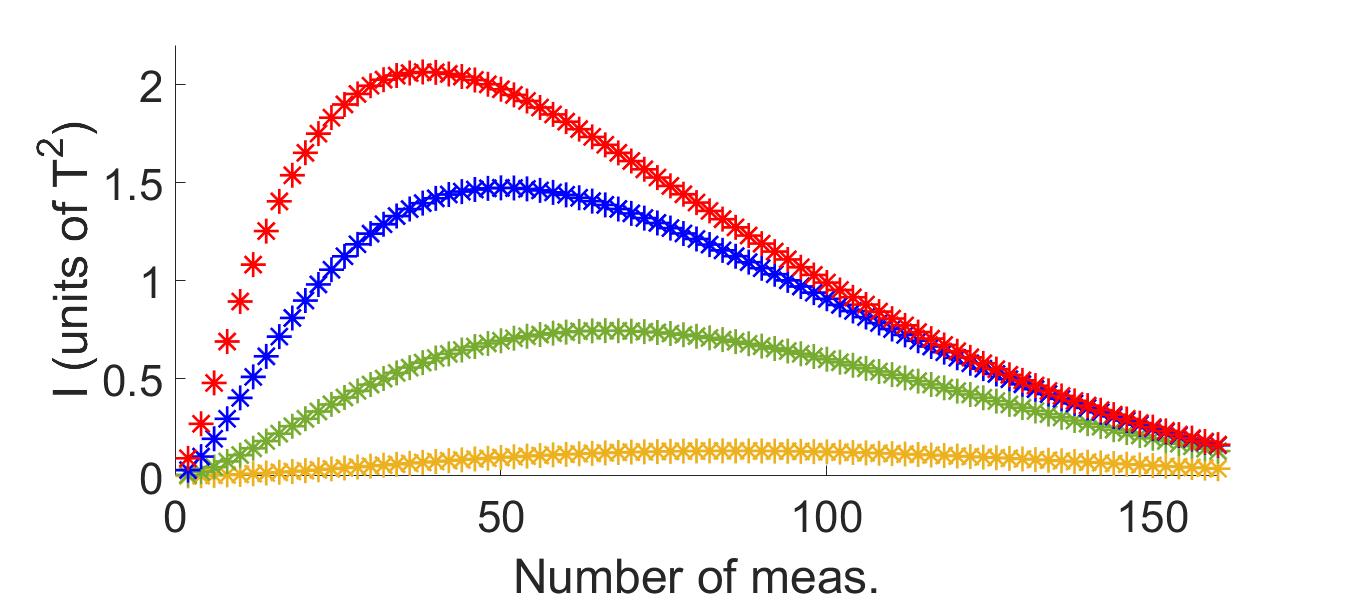}}
\subfigure[]{\includegraphics[width=7.2cm,height=2.7cm]{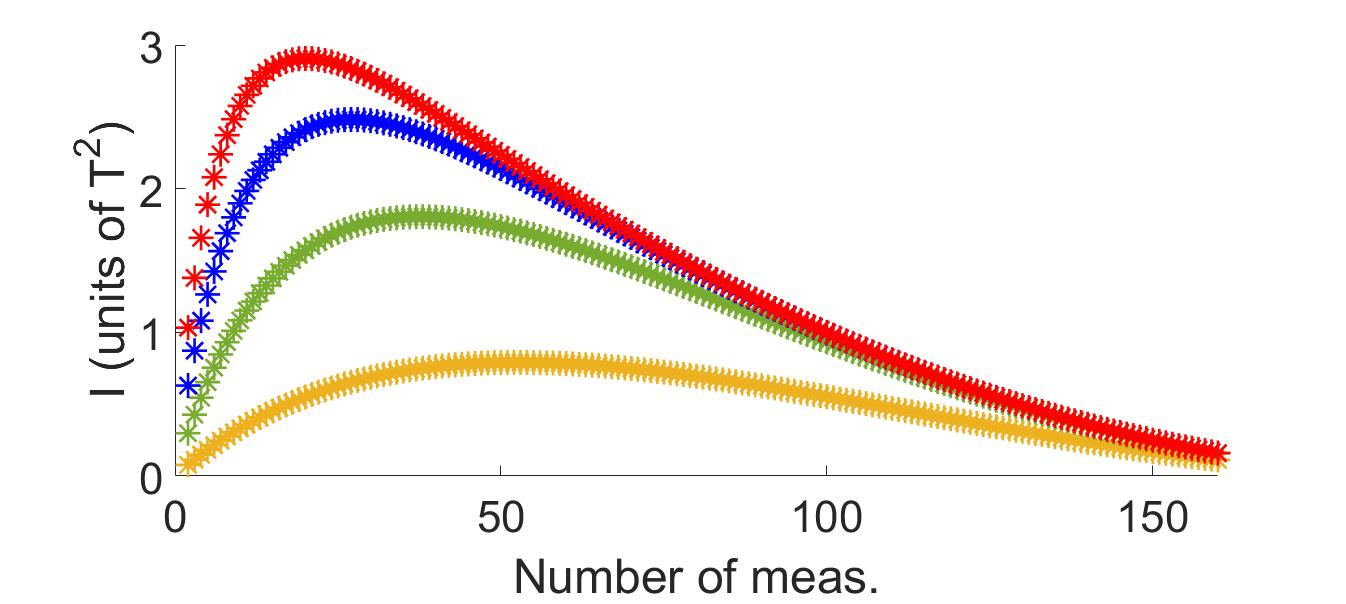}}
\subfigure[]{\includegraphics[width=7.4cm,height=2.9cm]{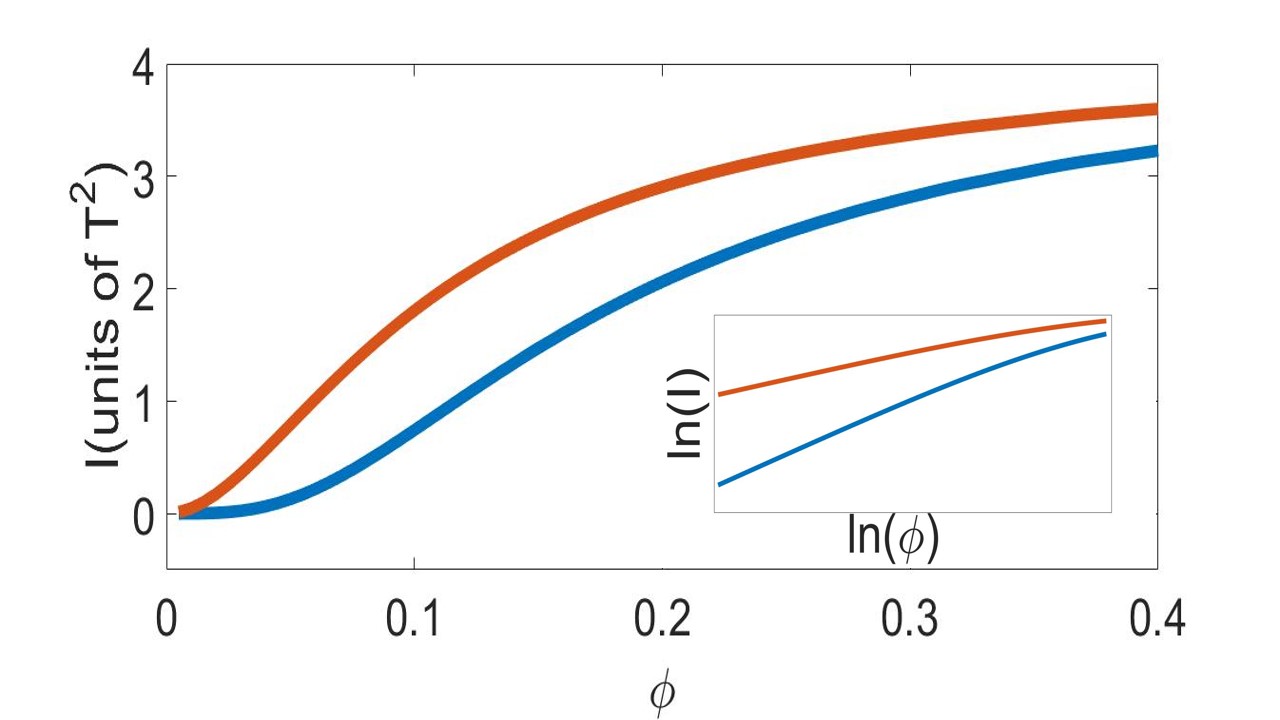}}
\subfigure[]{\includegraphics[width=7.2cm,height=2.7cm]{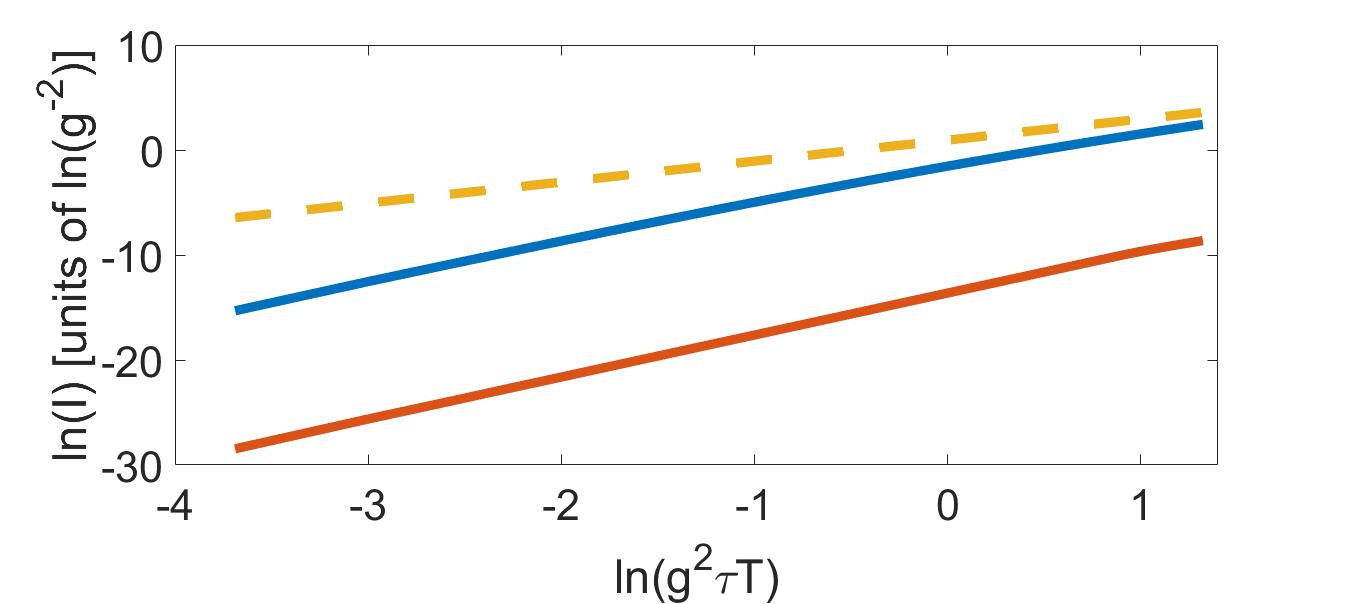}}\label{imperfect_fig}
\caption{FI analysis for multiple measurements at the detection periods. For all plots: $T$ denotes the duration of the experiment (limited by the coherence time of the nuclear spin), $\tau_{m}=0.005 [T],$ and $\phi$ is varied by changing $g.$ 
(a) FI as a function of total number of measurements, where the number of measurements at the beginning equals the number of measurements at the end. The different curves correspond to (from upper to lower): $\phi=0.2, \, 0.15, \, 0.1, \, 0.05 .$ Recall that the upper bound, achieved with strong measurements, equals to $4 T^2.$
(b) Same as (a), but now assuming a pure initial state (polarized), so measurement are applied only at the end (second detection period). The behavior is qualitatively the same as in (a), but better FI values are achieved.
(c) Comparison between the FI achieved with polarized and unpolarized nuclear spin (optimized over the number of measurements), for different values of $\phi.$ The upper (lower) curve corresponds to the polarized (unpolarized) case. The difference is significant for small values of $\phi,$ as for $\phi \ll 1$ the FI of polarized behaves as $\phi^{2},$ while the unpolarized as $\phi^{4}$ (as can be seen in the logarithmic plot and Eq. \ref{precision2}).
(d) FI (optimized over the number of measurements) as a function of time for the unpolarized case. The solid blue (upper) line represents the FI as a function of time, the dashed line is the ultimate limit ($4T^{2}$). For $g^{2}\tau T>1,$ the FI starts to converge to the ultimate limit. The solid red line (bottom) represents the FI with imperfect measurements ($a=0.05,b=0.7a$). 
Due to these imperfections, the convergence time to the ultimate limit is longer, and while far from this regime the FI achieved with these imperfections is much worse than with perfect measurements.}

\label{FI_analysis}
\end {center}
\end{figure}

Let us denote the total time of the experiment by $T,$ such that given a total number of measurements, $n,$ we have $T=\tau+n \tau_{m}.$ The Fisher information of a single experiment is bounded by $4 T^{2},$ which is achieved in a standard Ramsey experiment.
 This limit is obtained for large enough $g,$ such that $\phi=\frac{\pi}{4}$ is achievable,
when strong measurements can be applied. However in the weak coupling regime multiple measurements should be applied and finding the optimal number of measurements requires further investigation. Intuitively, increasing the number of weak measurements should provide better initialization and final detection. However, repeated measurements also reduce the accumulation of the phase (as the rotation period is shortened), thus there is a tradeoff between these two effects.  
Numerical values of the FI for different number of measurements are shown in fig. \ref{FI_analysis}. 
As expected, fig.   \ref{FI_analysis} illustrates that as the coupling strength is decreased, the optimal number of measurements increases and the FI decreases.

Note that at the limit of $\delta \ll g$ and $g \ll \omega,$ we can obtain a fairly good analytical approximation to the FI, by neglecting the detuning in the detection period, namely by neglecting $n\frac{\delta}{g},$ where $n$ is the total number of measurements and neglecting the interaction in the rotation period. This assumption is always correct towards the final stage of estimation as the uncertainty in the frequency is decreasing and the frequency of control($\omega_e$) can be adjusted in an adaptive manner.
Given these approximations the effective Hamiltonian in the detection period is just: $H_{eff}=g \sigma_{z} I_{X},$ and the effective Hamiltonian in the rotation period is just: $H_{eff}=\omega I_{Z},$ thus essentially we have a Ramsey experiment with weak initialization and weak final measurements. For convenience and simplicity, let us consider first the case in which the initial state of the nuclear spin is fully polarized (along the $X$ axis), such that only final weak measurements are required. 
It can be seen that for $n$ weak measurements we have $2^{n}$ possible trajectories, corresponding to $2^{n}$ possible measurement outcomes, however
the probability is determined only by the number of times in which we collapsed into $|\downarrow_{X}\rangle_{e}$ (and not by the order, which simplifies the calculation). 
Thus given that the nuclear spin is $|\uparrow_{X}\rangle_{n}$ and that the electron spin is initialized each time to $|\uparrow_{Y}\rangle_{e},$ the probability for collapsing $k$ times into $|\downarrow_{X}\rangle_{e},$ in a given order, reads:
\begin{eqnarray}
\begin{split}
& p_{k}=\sin\left(\phi+\pi/4\right)^{2k}\sin\left(\phi-\pi/4\right)^{2\left(n-k\right)}\sin\left(\omega\tau\right)^{2}+\\
&\sin\left(\phi-\pi/4\right)^{2k}\sin\left(\phi+\pi/4\right)^{2\left(n-k\right)}\cos\left(\omega\tau\right)^{2}.
\end{split}
\end{eqnarray}
Let us denote for convenience $c_{k}=\sin\left(\phi-\pi/4\right)^{2k}\sin\left(\phi+\pi/4\right)^{2\left(n-k\right)},$ and $d_{k}=\sin\left(\phi+\pi/4\right)^{2k}\sin\left(\phi-\pi/4\right)^{2\left(n-k\right)},$ with this notation:
 \begin{eqnarray}
 \begin{split}
 &p_{k}=c_{k}\cos\left(\omega\tau\right)^{2}+d_{k}\sin\left(\omega \tau \right)^{2}\\
 &= \frac{1}{2} \left(c_{k}+d_{k}\right)+ \frac{1}{2} \left(c_{k}-d_{k}\right)\cos\left(2\omega\tau\right).
 \end{split}
 \end{eqnarray}
Note that we can define $p_{\text{meas}}=\frac{c_{k}-d_{k}}{c_{k}+d_{k}},$ which is a natural way to quantify the strength of the measurement: $p_{\text{meas}}=\pm1$ corresponds to a trajectory in which the state collapsed. 
The FI for $n$ weak measurements is thus:
\begin{eqnarray}
\begin{split}
&I=\underset{k}{\sum}{n \choose k}\frac{\left(\frac{dp_{k}}{d\omega}\right)^{2}}{p_{k}}=\left(T-n\tau_{m}\right)^{2}\sin\left(2\omega\left(T-n\tau_{m}\right)\right)^{2}\times \\
& \underset{k}{\sum}{n \choose k}\frac{\left(c_{k}-d_{k}\right)^{2}}{c_{k}\cos\left(\omega\left(T-n\tau_{m}\right)\right)^{2}+d_{k}\sin\left(\omega\left(T-n\tau_{m}\right)\right)^{2}}.
\end{split}
\end{eqnarray}
The FI oscillates due to the factor of $\sin\left(2\omega\left(T-n\tau_{m}\right)\right)^{2},$ we can get rid of these oscillations and improve the FI by optimizing this phase. This can be done by changing the frequency of the pulses in the rotation period (and consequently changing $\omega$) or by  applying a pulse on the nuclear spin. 
Taking this phase to be $\frac{\pi}{4},$ we get:
\begin{eqnarray}
\begin{split}
I=2\left(T-n\tau_{m}\right)^{2}\underset{k}{\sum}{n \choose k}\frac{\left(c_{k}-d_{k}\right)^{2}}{c_{k}+d_{k}}\\
=4\langle p_{meas}^{2}\rangle\left(T-n\tau_{m}\right)^{2},
\label{FI_optimized}
\end{split}
\end{eqnarray}
where $\langle p_{meas}^{2}\rangle$ is the average of the measurement strength, $p_{\text{meas}}^{2},$ over all possible trajectories.
It can be seen that there is a trade-off between  $\langle p_{meas}^{2}\rangle,$ which monotonically increases with $n,$ and $\left(T-n\tau_{m}\right)^{2},$ which monotonically decreases with $n.$ In the limit of a single shot $\left ( \langle p_{meas}^{2}\rangle=1 \right)$  we recover the regular Ramsey limit.

Let us address now the case in which the nuclear spin is unpolarized, namely its density matrix is proportional to identity. Then in addition to the final weak measurements, one should apply weak measurements at the beginning in order to polarize the nuclear spin.
A very similar FI analysis applies to this case. We denote the number of measurements at the beginning (end) as $n_{1} \left( n_{2} \right),$ and $n=n_{1}+n_{2}$ stands for the total number of measurement. The probability now depends on the number of detections obtained at the beginning and at the end. The probability for a specific trajectory in which $|\uparrow_{X}\rangle_{e}$ is measured $k_{1} \left( k_{2} \right)$ times at the beginning (end) reads:
\begin{eqnarray}
\begin{split}
&p_{k_{1},k_{2}}= \frac{1}{2} \cos^{2}\left(\omega\tau\right)\left(c_{k_{1}}c_{k_{2}}+d_{k_{1}}d_{k_{2}}\right)+\\
&\frac{1}{2} \sin^{2}\left(\omega\tau\right)\left(c_{k_{1}}d_{k_{2}}+d_{k_{1}}c_{k_{2}}\right),
\end{split}
\end{eqnarray}
where we have used the same notation as above, namely: $c_{k_{i}}=\sin\left(\phi+\pi/4\right)^{2k_{i}} \sin \left(\phi+\pi/4\right)^{2\left(n_{i}-k_{i}\right)},$ and $d_{k_{i}}$ defined in an analogous manner. 
It would then be natural to define: $c_{k_{1,}k_{2}}=c_{k_{1}}c_{k_{2}}+d_{k_{1}}d_{k_{2}}
,$ and $d_{k_{1},k_{2}}=c_{k_{1}}d_{k_{2}}+d_{k_{1}}c_{k_{2}},$ such that the FI now reads:
\begin{eqnarray}
\begin{split}
&I=\left(T-N\tau\right)^{2}\underset{k_{1},k_{2}}{\sum}{n_{1} \choose k_{1}}{n_{2} \choose k_{2}}\frac{\left(c_{k_{1},k_{2}}-d_{k_{1},k_{2}}\right)^{2}}{c_{k_{1},k_{2}}+d_{k_{1},k_{2}}}\\
&=4\langle p_{\text{pol}}^{2}\rangle\langle p_{\text{meas}}^{2}\rangle\left(T-n\tau_{m}\right)^{2},
\label{FI_unpolarized}
\end{split}
\end{eqnarray}
where $p_{\text{meas}}=\frac{c_{k_{2}}-d_{k_{2}}}{c_{k_{2}}+d_{k_{2}}},$ as before, and $p_{\text{pol}}=\frac{c_{k_{1}}-d_{k_{1}}}{c_{k_{1}}+d_{k_{1}}}$ quantifies the strength of the polarization at the beginning.
The average is over all possible trajectories.
Note that this expression is symmetric with respect to $n_{1},n_{2}$ (first and second detection period), which implies that it reaches a maximum when $n_{1}=n_{2}=n/2.$
Numerical values of these expressions are presented in fig. \ref{FI_analysis}.

A natural question would be: Given $g,\tau_{m}$ and $T,$ what is the optimal number of measurements that should be applied, and what is the best achievable FI?
 To answer these questions, note that the time required for the weak measurements ($g \tau_{m} \ll 1$) to converge to a strong measurement goes as $\left(g^{2}\tau_{m}\right)^{-1}$ (see Appendix \ref{convergence_analysis} for a detailed analysis). Therefore the key quantity that determines the behavior of the FI is $g^{2}\tau_{m}T.$
For $T>\left(g^{2}\tau_{m}\right)^{-1},$ the FI starts to converge to the ultimate limit of $4T^{2},$  and the optimal number of measurements goes as $\left(g\tau_{m}\right)^{-2}$. 
Far from the strong measurement regime, for $T\ll\left(g^{2}\tau_{m}\right)^{-1},$ we have that $\langle p_{\text{pol}}^{2}\rangle,\langle p_{\text{meas}}^{2}\rangle \ll 1,$ and they grow linearly with the number of measurements. Therefore, for the unpolarized case, the optimum would be to dedicate roughly half of the time for measurements, which yields: $I \sim g^{4}\tau_{m}^{2}T^{4}.$
For the polarized case, the optimum would be to dedicate roughly one third of the time to measurements, which yields: $I \sim g^{2}\tau_{m} T^{3}$ (note that this behavior coincides with that of a classical signal \cite{schmitt2017submillihertz}). 

Observe that in the weak measurement regime ($g^{2}\tau_{m}T\ll1$), the polarized and unpolarized cases differ by a factor of $g^{2}\tau_{m}T,$ which is quite significant.
While in the strong measurement regime ($g^{2}\tau_{m}T\gg1$) the FI of both the polarized and the unpolarized converge to the same value of $4T^2$.

 It is interesting to compare this method to the NMR method, which has been recently used for sensing classical signals\cite{schmitt2017submillihertz,bucher2017high,boss2017quantum,rotem2017limits}. In the NMR method, many synchronized consecutive measurements are performed, where the detuning is kept fixed throughout the entire experiment \footnote{In general,  there can be a time interval between two consecutive measurements, in which the nuclear spin will evolve freely. In our analysis we assume there are no such intervals between the measurements. }.
The outcomes of these measurements then undergo a Fourier-transform analysis or a Maximum-likelihood estimation to obtain the frequencies of the signal. When dealing with classical signals there is no backaction and thus performing multiple measurements all along the experiment is the optimal thing to do.
We analyzed the FI achieved with the NMR method for sensing a single nuclear spin, in which backaction effects cannot be neglected, the detailed analysis presented in appendix \ref{NMR method}.
Just like with our method, the polarized and unpolarized cases coincide for $g^{2}\tau_{m}T\gg1.$ However they do not converge to the ultimate limit of of $4T^{2},$ but rather to a scaling of $T,$ which is of course much worse.
The reason for this behavior can be simply understood: The weak measurements of the nuclear spin extract information about its state but also perturb it. The number of weak measurements required to polarize and measure the nuclear spin strongly goes as $\left(g\tau_{m}\right)^{-2}.$ Therefore if $T\gg\left(g^{2}\tau_{m}\right)^{-1},$ applying measurements all throughout the experiment will result in destroying the oscillations without gaining any further information.    
In other words, measurements are needed to extract information, but with too many measurements the backaction induces decoherence that affects the FI.

\begin{figure}
\begin {center}
\includegraphics[width=9cm]{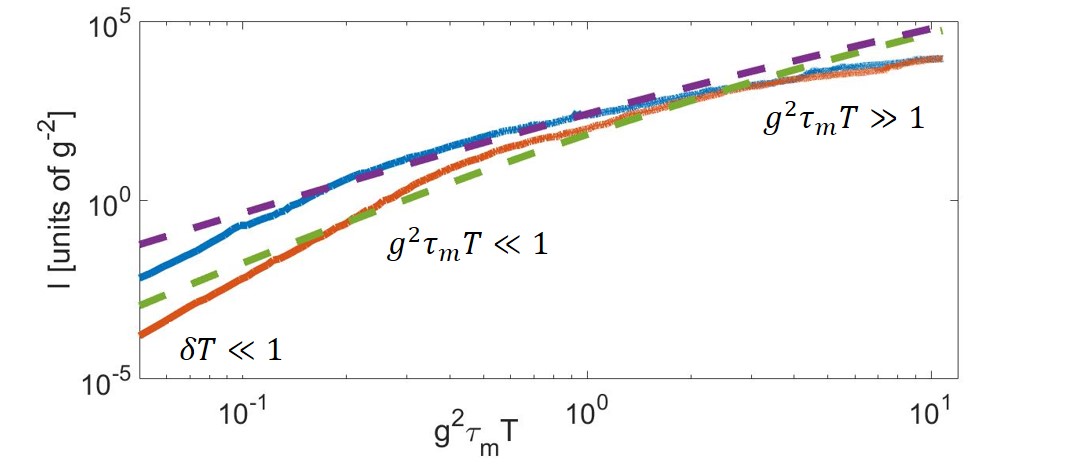}
\caption{A comparison between the FI achieved with the NMR method (solid curves) and our method (dashed curves). The upper curves (dashed and solid) correspond to the polarized scenario and the lower curves to the unpolarized. 
In both methods, the polarized and the unpolarized curves coincide after long enough time ($g^2 \tau_{m}T \gg 1$). While the FI with NMR method converges to a scaling of $T,$ the FI with our method converges to the ultimate limit of $4T^{2}.$      }

\label{nmr_main_text}
\end {center}
\end{figure}

It can be also observed that as $\delta T\rightarrow0$ the FI vanishes, just because the oscillations of the probability start from the antinode, in which the derivative with respect to $\delta$ vanishes. 
In the regime of $\delta T>1$ and $T \ll \left(g^{2}\tau_{m}\right)^{-1},$ i.e. weak measurement regime (accumulative effect of the backaction is small), the performance of the NMR method coincides with our method. In fact, in this regime the NMR method even outperforms our method (typically by a factor of 2), as can be observed in fig. \ref{nmr_main_text}. However, as already discussed, for $T \gg \left(g^{2}\tau_{m}\right)^{-1},$ in which the weak measurements start to converge to a strong measurements, our method outperforms the NMR. A comparison between both methods is shown in fig. \ref{nmr_main_text}.

{\it {Imperfect measurements}}---In all the analyses described above, we have assumed perfect quantum measurements. However, in NV centers, as well as in some other experimental platforms, measurements suffer from severe limitations. In NV centers, the information obtained from a measurement is the number of emitted photons. 
The problem is that photons are emitted from both the bright and the dark states, both states collapse to the same state (not a QND measurement), and the photodetection efficiency is poor \footnote{QND measurements of the electron spin in NV center can be implemented, by entangling it to the nuclear spin of the Nitrogen. This however comes at the expense of time, these measurements are much slower.}.
This results in a very low efficiency. It is therefore necessary to compare the FI achieved with these limitations to the original FI.

The measurement can be modeled as a Bernouli trial (photon detected/not detected) with a probability of $p=ap_{b}+bp_{d},$ where $p_{b} \, (p_{d})$ is the probability of the bright (dark) state, and $a \, (b)$ represents the probability to get a photon from the bright (dark) state.{\footnote{We assume here a low photodetection efficiency, otherwise it is possible to get more photons and the analysis should be modified.}}
Applying the protocol of two weak measurements, one at the beginning and one at the end.Then, with the imperfections considered above, the resulting FI reads:
\begin{eqnarray}
\begin{split}
&I=\left[\frac{2\left(a-b\right)^{2}}{\left(a+b\right)\left(2-a-b\right)}\sin\left(2g\tau_{m}\right)^{2}\right]^{2}T^{2}\\
&\approx16\frac{\left(a-b\right)^{4}}{\left(a+b\right)^{2}}\phi^{4}T^{2}.
\label{two_imperfect_measurements}
\end{split}
\end{eqnarray}  
Note that this result differs from the result obtained with perfect measurements (eq. \ref{precision2}) by a factor of $\left(\frac{\left(a-b\right)^{2}}{2 \left(a+b \right)}\right)^{2}.$
For typical experimental values of $a,b$ (for example $a=0.05,\:\frac{b}{a}=0.7$), this yields a difference of six orders of magnitude.

Employing multiple measurements can reduce this difference.  The FI with imperfect measurements still converges to $4 T^{2},$ given a long enough coherence time of the nuclear spin.
However, the time required for this convergence is roughly $\left(\frac{\left(a-b\right)^{2}}{a+b}g^{2}\tau_{m}\right)^{-1}$ (see appendices \ref{Imperfect measurements},\ref{convergence_analysis}), which is much longer than the time required with perfect measurements ($\left(g^{2}\tau_{m}\right)^{-1}$).  
Again, considering characteristic experimental values (for example $g\tau_{m}=0.1$ and same $a,b$ as before), the imperfect measurements add two orders of magnitude $(T\sim 10^{4} \tau_{m})$ to the time required to reach the strong measurement regime.
Given that we are far away from the strong measurement regime, i.e. $\frac{\left(a-b\right)^{2}}{a+b}g^{2}\tau_{m}T\ll1$, the optimal number of measurements, for the unpolarized case, would be $N/2$ (where $N=\frac{T}{\tau_{m}}$) and as a result $I\sim\frac{\left(a-b\right)^{4}}{\left(a+b\right)^{2}}g^{4}\tau_{m}^{2}T^{4}.$
A comparison between the FI achieved with perfect and imperfect measurements is presented in fig. \ref{FI_analysis}.
More details on the calculations of the FI for imperfect measurements are found in appendix \ref{Imperfect measurements}. 

Note that since in the weak measurement regime there is not much difference between the performance of this method and the NMR method, and since due to the imperfect measurements almost any reasonable experimental realization is in the weak measurement regime, then in practice this method and the NMR method should have similar performance.             

\subsection{Effects of classical noise}  
In this section we address  our ability to differentiate between the quantum signal and classical noise.
The resolution that was previously described will be significantly reduced by classical noise that might have a considerable power spectrum at the Larmor frequency.
The quantum signal is described by Eq.\ref{main} while the classical signal/noise can be described by
$H_{C1} = g\sigma_z \left( \cos(\omega_n t +\varphi)   \right),$ where $\varphi$ and $g$ are random variables which are uniformly distributed.
The frequency $\omega_n$ should also be a random variable, but we will assume that $\omega_n$ is identical to the frequency of the nuclei to make it harder to distinguish.
In the following we present a few methods to differentiate between the two signals.

As the role of the series of measurement at the beginning is to polarize the nuclei, it is instructive to see what would be the role of polarization in differentiating between the signal and the noise.
The nucleus can be polarized at the beginning using dynamical nuclear polarization\cite{abragam1978principles,london2013detecting,goodman1972spin,PhysRev.128.2042}.
In that case the final measurement result gives the average: 
\begin{equation}
p_{\uparrow_x} - p_{\downarrow_x} = 2 p_{\text{pol}} \sin \left(2 \phi \right) \cos (2 \omega  \tau ),
\label{Pol_HH}
\end{equation}
where $p_{\text{pol}}$ is the parameter that characterizes the polarization of the nuclear spin, namely: $p_{\text{pol}}=0.5\left(p_{|\downarrow_{x}\rangle_{n}}-p_{|\uparrow_{x}\rangle_{n}}\right).$
Regarding the classical signal/noise, one can observe that after averaging over the random phase $\varphi,$ we have that $p_{\uparrow_x} =p_{\downarrow_x},$ which allows us to differentiate between the classical and the quantum signals.   The classical noise, however, decreases the coherence time of the NV which limits the signal to noise ratio.

As it is clear that polarization makes a clear distinction between the classical and the quantum signal, it is natural to ask whether a similar distinction can be made when the nuclear spin is unpolarized.
As already shown, we can induce polarizations by applying a measurement, but the polarization direction would be random. 
Therefore an extra ingredient is needed: The signals can be differentiated by operating directly on the nuclei. The first measurement creates a small polarization in either the $x$ or $-x$ direction.
By adding a $\pi$ pulse on the nuclear spin in one of these cases a net polarization is created: $p_{\text{pol}}=\sin\left(2\phi\right)/2$. Thus in the second measurement we get
\begin{equation}
p_{\uparrow_x} - p_{\downarrow_x} =  \sin^2 \left( 2 \phi \right) \cos (2 \omega  \tau ).
\label{Pol_Meas}
\end{equation}
In the classical case however, the $\pi$-pulse has no effect, and thus the second measurement will results in a $50\%$ up,  $50\%$ down result. Thus the classical noise will average to zero and the probability of getting a $\ket \uparrow_x$
result will not change due to the classical noise. However, also in this case the coherence time of the NV will be limited by the noise. The difference between Eq. \ref{Pol_Meas} and Eq. \ref{Pol_HH} is due to the fact that the initial  polarizarion is proportional to  $\sin \left( 2 \phi \right)$ which creates a different dependance on $\phi$.

Instead by operating directly on the nuclei it is also possible to induce the operation via the NV, by realizing the same Hamiltonian (Eq. \ref{main}), as by using this interaction it is possible to rotate the nuclei and induce a $\pi$(or less) flip.
Thus, the sequence will be the following: the first measurement creates a small polarization in either the $x$ or $-x$ direction.
In the later case a rotation of phase $\phi_1$ is realized via Hamiltonian (Eq. \ref{main}) around the $x$ axis and thus a small polarization is created.
In that case the correlation functions is:
\begin{equation}
p_{\uparrow_x} - p_{\downarrow_x}  \approx 2 \phi ^2 \sin ^2\left(\frac{\phi_1}{2}\right) \cos (2 \omega  \tau ),
\end{equation}
while the classical noise is averaged to zero as in the previous cases.

{\it Conclusion and outlook ---} This manuscript presents a quantum sensing scheme 
for detecting the frequencies of quantum systems. Analysis of the scheme shows that the resolution is limited mainly by the coherence time of the target system.
While the coherence time of the probe affects only the strength of the initial polarization and the final measurement. 
A Fisher information (FI) analysis of the scheme was presented, and it was shown that for long enough coherence time of the nuclear spin the FI converges to the ultimate limit.  

The performance of this protocol was compared to that of the NMR method, which was used for sensing classical signals\cite{schmitt2017submillihertz,bucher2017high,boss2017quantum,rotem2017limits}.
The performance of our method is similar to that of the NMR method in the weak measurement regime, and significantly outperforms it in the the strong measurement regime.
It should be noted that the NMR method can be further analyzed, in fact finding analytical expression of the FI of this method is an open challenge. 
 It would be also interesting to further inquire the performance of both methods for sensing multiple target spins.

{\it Acknowledgements ---} We thank Amit Rotem, Daniel Louzon and Nati Aharon for fruitful discussions. A. R. acknowledges the support of the Israel Science Foundation(grant no. 039-8823), the support of the European commission (EU Project DIADEMS). T.G. is supported by the Adams Fellowship Program of the Israel Academy of Sciences and Humanities. M.K. is supported by the Israel Science
Foundation, Grant No. 1287/15.

\appendix
\section{Imperfect measurements}
\label{Imperfect measurements}
We now consider the realistic scenario in which the projective measurements cannot be achieved, as there is a finite detection efficiency, a finite probability to get photons from the dark state, and both the dark and bright states collapse to the same state (it is not a QND measurement).
This issue was discussed in the main text, we present here the calculations in more details.
As in the main text, let us denote the probability to get a photon from the bright state ($|\downarrow_{y}\rangle_{e}$) as $a,$ and $b$-from the dark state. Note that the information we get from each measurement is whether or not a photon was emitted.
We can make a similar analysis of the FI, using the same approximations as in the main text.
 Let us start again with the simpler case in which the nuclear spin is initially polarized. Denoting $\alpha=\sin\left(g\tau_{m}+\pi/4\right)^{2},\:\beta=\sin\left(g\tau_{m}-\pi/4\right)^{2},$ then the probability for a detection of a photon in a single measurement reads:
\begin{equation}
\cos^{2}\left(\omega\tau\right)\left(a\alpha+b\beta\right)+\sin^{2}\left(\omega\tau\right)\left(a\beta+b\alpha\right),
\end{equation}
and the probability that no photon was detected reads:
\begin{eqnarray}
\begin{split}
&\cos^{2}\left(\omega\tau\right)\left(\left(1-a\right)\alpha+\left(1-b\right)\beta\right)\\
&+\sin^{2}\left(\omega\tau\right)\left(\left(1-a\right)\beta+\left(1-b\right)\alpha\right).
\end{split}
\end{eqnarray}
It can now be observed that given $n$ measurements, the probability for a specific trajectory in which $k$ photons are emitted is:
\begin{eqnarray}
\begin{split}
&\cos^{2}\left(\omega\tau\right)\left(a\alpha+b\beta\right)^{k}\left(\left(1-a\right)\alpha+\left(1-b\right)\beta\right)^{n-k}+\\		
&\sin^{2}\left(\omega\tau\right)\left(a\beta+b\alpha\right)^{k}\left(\left(1-a\right)\beta+\left(1-b\right)\alpha\right)^{n-k}.
\end{split}		
\end{eqnarray}
Using the same notation as in the main text, we have:
\begin{eqnarray}
\begin{split}
&c_{k}=	\left(a\alpha+b\beta\right)^{k}\left(\left(1-a\right)\alpha+\left(1-b\right)\beta\right)^{n-k}\\
&d_{k}=\left(a\beta+b\alpha\right)^{k}\left(\left(1-a\right)\beta+\left(1-b\right)\alpha\right)^{n-k}.
\label{general_c_d}
\end{split}
\end{eqnarray}
We can define $p_{\text{meas}}=\frac{c_{k}-d_{k}}{c_{k}+d_{k}},$ just like in the main text, and then the FI has the same form as in eq. \ref{FI_optimized}.

\begin{figure}[h]
\begin {center}
\subfigure[]{\includegraphics[width=7cm]{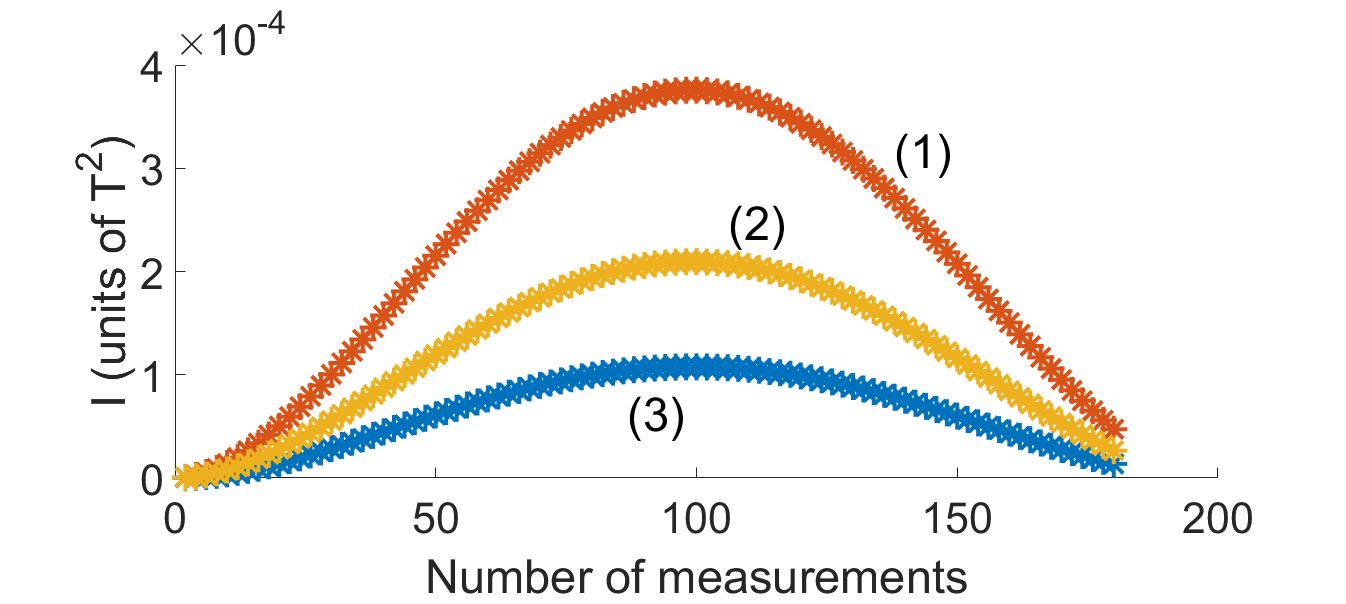}}
\subfigure[]{\includegraphics[width=7cm]{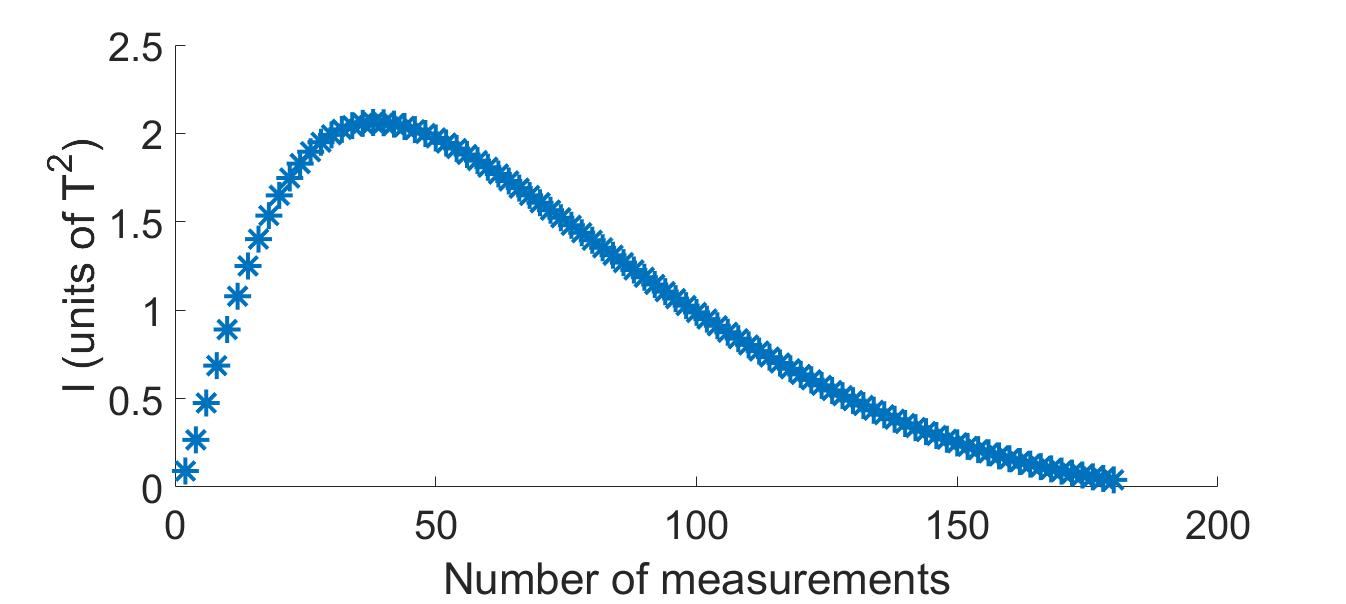}}
\subfigure[]{\includegraphics[width=6cm]{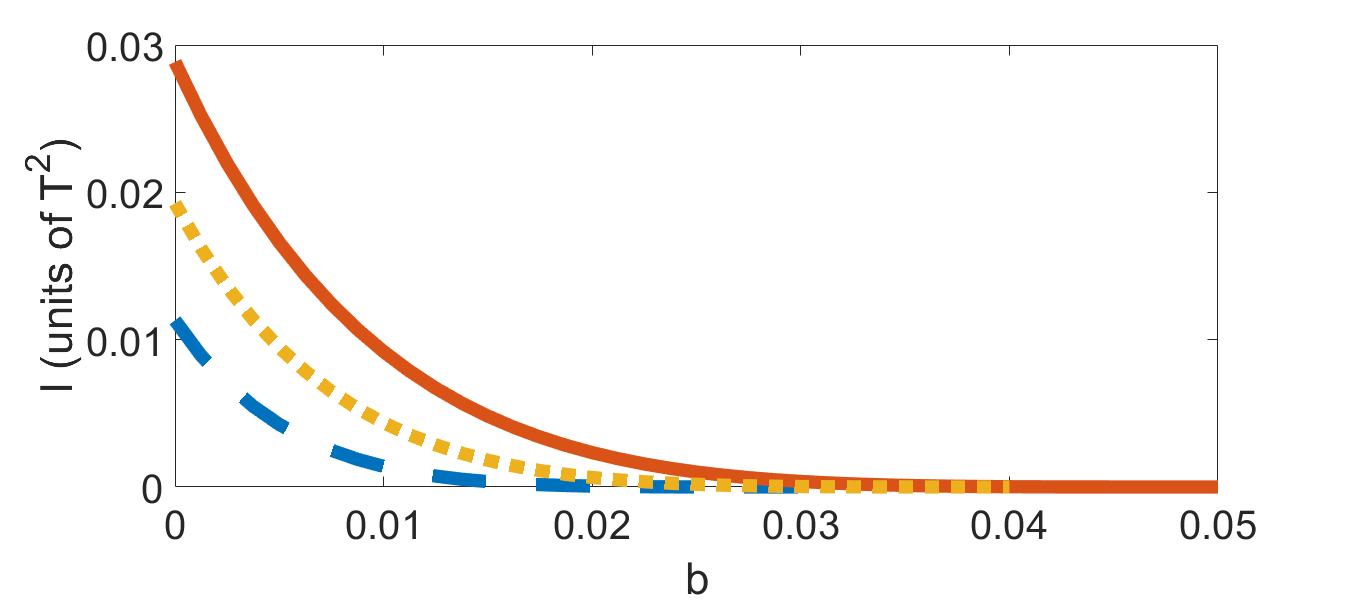}}
\caption{ FI as a function $a$ and $b,$ where $a$ is the probability to detect photons from the bright state, and $b$ is the probability to detect photons from the dark state. The optimal case is $a=1,b=0,$ which corresponds to perfect measurements.
(a) The FI as a function of the number of measurement for different values of $b.$ In this figure $a=0.05,$ and (1),(2),(3) correspond to contrasts of $0.4,0.35,0.3$ (namely $b=0.6a,0.65a,0.7a$)  respectively. These results can be compared to the case of perfect measurement, which is presented in panel (b).
Observe that the for the imperfect case, the number of measurements is much larger, and about half of the possible time is spent over measurements. There is a difference of 4 orders of magnitude between the perfect and the imperfect scenarios, meaning that for these imperfect measurements $10^4$ experiments are required in order to get a precision of $1/T.$  
(c) The optimal FI (optimized over the number of measurements), as a function of $b,$ for different values of $a$: $a=0.05\left(\text{upper curve}\right),0.04\left(\text{middle curve}\right),0.03\left(\text{lower curve}\right).$ Clearly the optimal FI drops as $a$ becomes smaller and as the contrast between $a$ and $b$ becomes smaller.
For all plots: $g=40\:\left[T^{-1}\right],\tau_{m}=5\cdot10^{-3}\:[T],$ meaning the ratio between the $T_{2}$ and the coherence time of the nuclear spin was taken to be 200. }
\label{FI_analysis_imperfect}
\end{center}
\end{figure}

Regarding the unpolarized case, in which $n_{1}\,\left(n_{2}\right)$ measurements are performed in the first (second) detection period, then the probability for a trajectory in which $k_{1}\:\left(k_{2}\right)$
photons are detected reads: 
\begin{eqnarray}
\begin{split} 
0.5\cos^{2}\left(\omega\tau\right)\left[c_{k_{1}}c_{k_{2}}+d_{k_{1}}d_{k_{2}}\right]+\\
0.5\sin^{2}\left(\omega\tau\right)\left[c_{k_{1}}d_{k_{2}}+d_{k_{1}}c_{k_{2}}\right]		
\end{split}
\end{eqnarray}
and then we have:
\begin{eqnarray}
\begin{split} 
&c_{k_{1},k_{2}}=0.5\left[c_{k_{1}}c_{k_{2}}+d_{k_{1}}d_{k_{2}}\right]\\
&d_{k_{1},k_{2}}=0.5\left[c_{k_{1}}d_{k_{2}}+d_{k_{1}}c_{k_{2}}\right].
\end{split}
\end{eqnarray}
So the FI has the same form as in eq. \ref{FI_unpolarized}. Numerical results are presented in fig. \ref{FI_analysis_imperfect}. 

Due to these imperfections, $\langle p_{\text{meas}}^{2}\rangle$ is smaller implying that the optimal number of measurements is increased and the FI drops.
It is shown in appendix \ref{convergence_analysis} that with imperfect measurements the time required for the FI to converge to the ultimate limit goes as $\frac{a+b}{\left(a-b\right)^{2}}\left(g^{2}\tau_{m}\right)^{-1},$ namely an additional prefactor of $\frac{a+b}{\left(a-b\right)^{2}}$ compared to perfect measurements.    

In the numerical results, realistic values of $a$ and $b$ were taken ($0.005$ and $0.0035$), and the FI has been decreased by four orders of magnitude   
More concretely, without these imperfection the precision is about $1/T,$ whereas with these imperfections one needs to perform $10^4$ independent experiments to reach the same level of precision.

\section{Convergence time to strong Ramsey regime}
\label{convergence_analysis}
\begin{figure}
\begin {center}
\includegraphics[width=9cm,height=6cm]{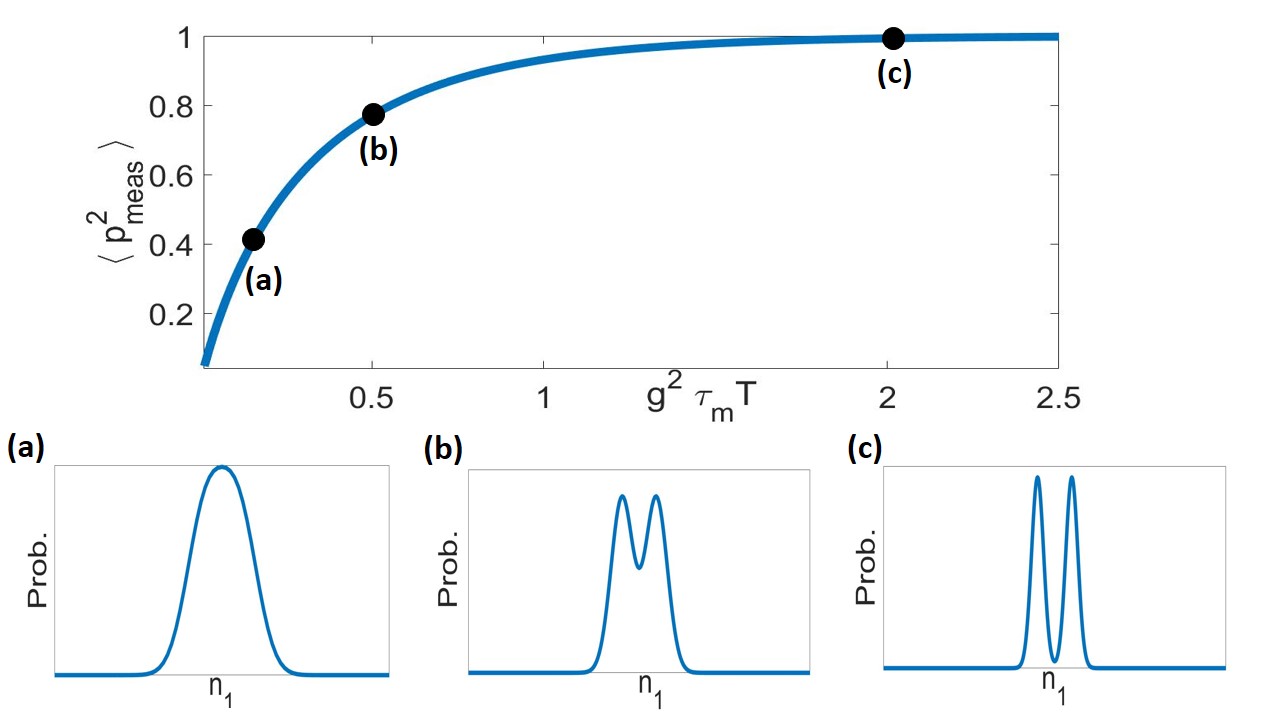}
\caption{ Convergence time to a strong measurement: The scheme converges to a standard Ramsey measurement as the weak measurements converge to a strong measurement, quantified by $\langle p_{\text{meas}}^{2}\rangle\rightarrow1.$
In this plot the convergence of $\langle p_{\text{meas}}^{2}\rangle$ is illustrated (for perfect measurements). The convergence time goes as $\left(g^{2}\tau_{m}\right)^{-1}.$
The distribution of the outcomes is the sum of two binomial distributions and the convergence time is the time required for these two distributions to be well separated, as is illustrated in this plot. 
For imperfect measurements the convergence time goes as $\frac{a+b}{\left(a-b\right)^{2}}\left(g^{2}\tau_{m}\right)^{-1}.$   }

\label{convergence_to_strong}
\end {center}
\end{figure}

In this section we seek to derive the convergence time of $\langle p_{\text{meas}}^{2}\rangle$ (and $\langle p_{\text{pol}}^{2}\rangle$) to $1.$
Note that: 
\begin{equation}
\langle p_{\text{meas}}^{2}\rangle=0.5\underset{k}{\sum}{n \choose k}\frac{\left(c_{k}-d_{k}\right)^{2}}{c_{k}+d_{k}},
\end{equation}
where the general $c_{k},d_{k}$ are given by eq. \ref{general_c_d}.
Observe that the probability to get $k$ photons in this sequence of weak measurements is given by: $p\left(k\right)=0.5\left(p_{1}\left(k\right)+p_{2}\left(k\right)\right),$ where $p_{1}, p_{2}$ are binomial distributions given by:
\begin{eqnarray}
\begin{split}
&p_{1}\left(k\right)={n \choose k}c_{k}={n \choose k}\left(a\alpha+b\beta\right)^{k}\left(\left(1-a\right)\alpha+\left(1-b\right)\beta\right)^{n-k} \\
&p_{2}\left(k\right)={n \choose k}d_{k}={n \choose k}\left(a\beta+b\alpha\right)^{k}\left(\left(1-a\right)\beta+\left(1-b\right)\alpha\right)^{n-k}.
\end{split}
\end{eqnarray}
It is now simple to see that $\langle p_{\text{meas}}^{2}\rangle$ can be also written as: 
\begin{equation}
\langle p_{\text{meas}}^{2}\rangle=0.5\underset{k}{\sum}\frac{\left(p_{1}\left(k\right)-p_{2}\left(k\right)\right)^{2}}{p_{1}\left(k\right)+p_{2}\left(k\right)},
\end{equation}    
which implies:
\begin{equation}
\langle p_{\text{meas}}^{2}\rangle=0.5\underset{k}{\sum}\frac{\left(p_{1}\left(k\right)-p_{2}\left(k\right)\right)^{2}}{p_{1}\left(k\right)+p_{2}\left(k\right)}\leq\underset{k}{\sum}0.5\left(p_{1}\left(k\right)+p_{2}\left(k\right)\right)=1.
\end{equation}    
The inequality is saturated if and only if $p_{1}$ and $p_{2}$ are well separated (no overlap between the distributions). Note that these Binomial distributions become separated if the difference between the mean values is smaller than the sum of the standard deviations.This basically sets the condition for convergence and thus determines the convergence time.
 Hence this sequence of measurements converges to a strong measurement when the distribution of the outcomes consists of two well separated Binomial distributions, as is illustrated in Fig. \ref{convergence_to_strong}.
  
 Let us find this time for perfect and imperfect measurements. For perfect measurements the probabilities of the Binomial distributions are $p=\cos^{2}\left(g\tau_{m}+\pi/4\right),\: 1-p=\cos^{2}\left(g\tau_{m}-\pi/4\right),$ which means that the condition is: $\frac{2\sqrt{p\left(1-p\right)}}{\sqrt{N}}\leq1-2p.$ Inserting the expressions for $p,$ we have that: $N\geq\cot^{2}\left(2g\tau_{m}\right).$ Hence for $g\tau_{m}\ll1,$ we have that $N\geq\frac{1}{4\left(g\tau_{m}\right)^{2}},$ which implies that the convergence time goes as:
 \begin{equation}
T \sim \left(g^{2}\tau_{m}\right)^{-1}.  
\end{equation}
For imperfect measurements the probabilities of the Binomial distributions are: $p_{1}=a\alpha+b\beta,\:p_{2}=a\beta+b\alpha,$ and the condition is then $\sqrt{\frac{p_{1}\left(1-p_{1}\right)}{N}}+\sqrt{\frac{p_{2}\left(1-p_{2}\right)}{N}}\leq p_{2}-p_{1}.$ 
For $a,b \ll 1$ (the realistic case) we get that the condition reads: $N\geq\frac{a+b+2\sqrt{\left(a+b\right)^{2}-\sin\left(2g\tau_{m}\right)^{2}\left(a-b\right)^{2}}}{\left(a-b\right)^{2}\sin\left(2g\tau_{m}\right)^{2}}.$ Assuming weak measurements ($g\tau_{m} \ll 1$) we get that: $N\geq\frac{3\left(a+b\right)}{4\left(a-b\right)^{2}\left(g\tau_{m}\right)^{2}}.$ Therefore the time required for convergence goes as: 
\begin{equation}
T\sim\frac{a+b}{\left(a-b\right)^{2}}\left(g^{2}\tau_{m}\right)^{-1}.   
\end{equation}

\section{Comparison with NMR technique: Consecutive measurements}
\label{NMR method}
The method we presented in this manuscript consists of detection periods, in which $\delta$ is taken to be as small as possible and measurements are applied, as well as a rotation period in which the nuclear spin rotates freely with Larmor frequency (and no measurements are applied).
Current nano NMR protocols, however, usually employ a different method. A typical nano NMR experiment consists of many consecutive measurements, where $\delta$ is kept fixed throughout the experiment. This nano NMR method is exactly what is used to sense classical signals \cite{schmitt2017submillihertz,bucher2017high,boss2017quantum}, and an ensemble of nuclear spins.
It would be instructive to compare between these two methods when applied to sense a frequency of a single spin.

Our analysis of this method assumes that measurements are applied one immediately after the other. In principle, one can set a break between measurements, in which the nuclear spin evolves freely. 
This time interval can be modified and optimized. We will address this point later.  

Let us analyze the nano NMR protocol. In each measurement step, the operators acting on the nuclear spin (after tracing out the electron spin degrees of freedom) are:
\begin{eqnarray}
\begin{split}
&C_{\pm}=\frac{1}{\sqrt{2}} [\cos\left(\sqrt{g^{2}+\delta^{2}}\tau_{m}\right)-i\sin\left(\sqrt{g^{2}+\delta^{2}}\tau_{m}\right)\frac{\delta}{\sqrt{g^{2}+\delta^{2}}}I_{Z}\\
&\pm\sin\left(\sqrt{g^{2}+\delta^{2}}\tau_{m}\right)\frac{g}{\sqrt{g^{2}+\delta^{2}}}I_{X}],
\end{split}
\end{eqnarray}
 (followed by a normalization), where $C_{+}\:\left(C_{-}\right)$
acts on the nuclear spin with a probability of $\text{Tr}\left(C_{+}\rho C_{+}^{\dagger}\right)$$\left(\text{Tr}\left(C_{-}\rho C_{-}^{\dagger}\right)\right).$
Since we are interested in the limit of $g\tau_{m},\delta\tau_{m}\ll1,$ these operators
can be approximated as: 
\begin{eqnarray}
\begin{split}
&C_{\pm}\approx\frac{1}{\sqrt{2}}\left[I+i\delta\tau I_{Z}\pm g\tau I_{X}\right]\\
&\approx\left(\cos\left(g\tau_{m}\right)\pm\sin\left(g\tau_{m}\right)I_{X}\right)\left(\cos\left(\delta\tau_{m}\right)-i\sin\left(\delta\tau_{m}\right)I_{Z}\right),
\end{split}
\end{eqnarray}
namely in leading order of $g\tau_{m},\delta\tau_{m}$ these operators are
just a rotation by an angle of $\delta\tau_{m}$ followed by a weak measurement
of $I_{X}$ (with a strength of $\phi=g\tau_{m}$).

\begin{figure}
\begin {center}
\subfigure[]{\includegraphics[width=6.5cm]{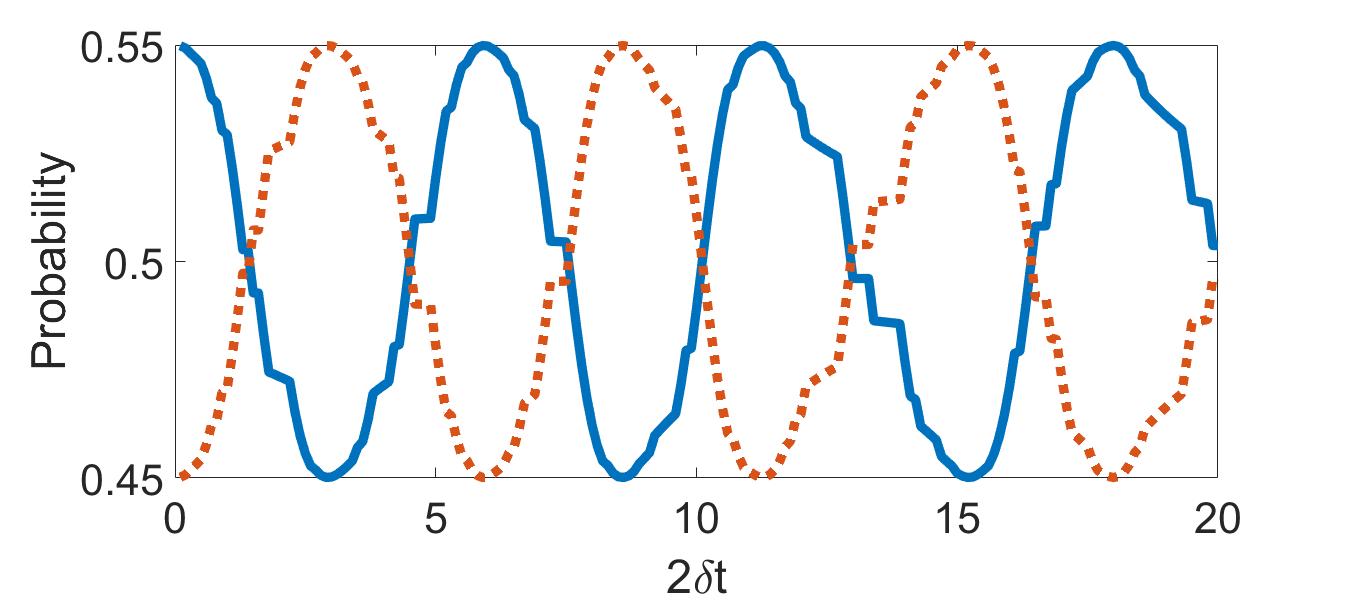}}
\subfigure[]{\includegraphics[width=6.5cm]{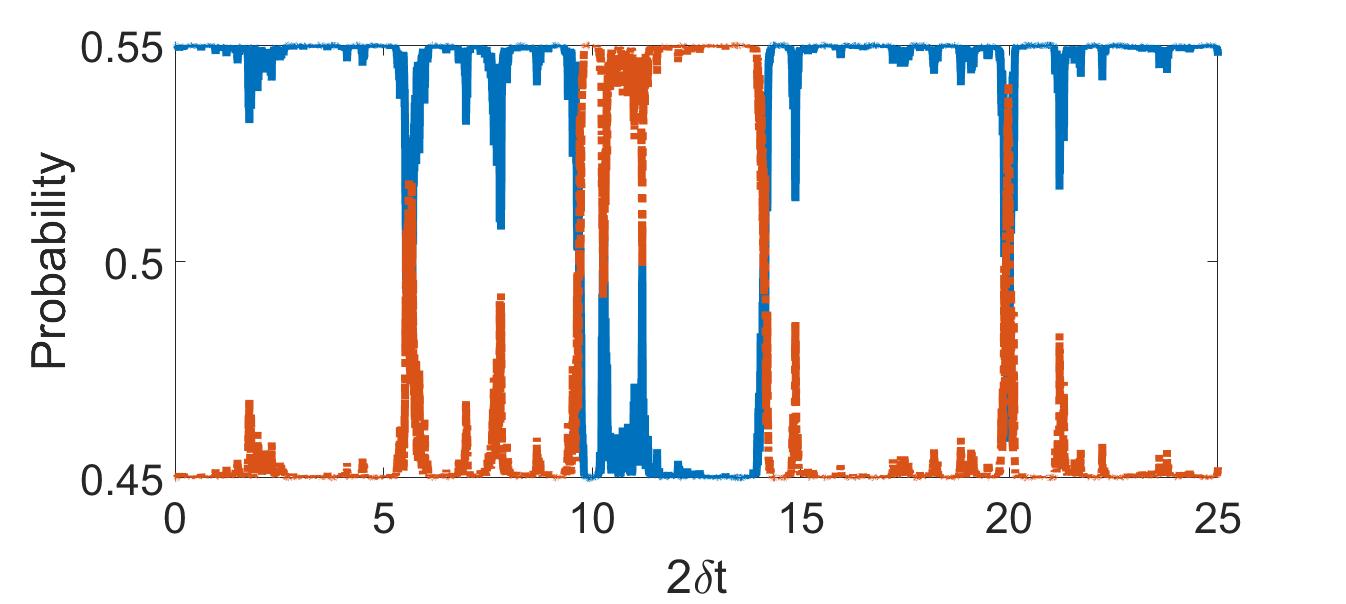}}
\caption{Dynamics of a polarized spin as a function of time for a given coupling strength and different detunings in the NMR method. The blue (solid) and the red (dotted) curves correspond to $p_{+}, p_{-}$ (the collapse probabilities of the electron spin on $|\uparrow_{y}\rangle_{e}/|\downarrow_{y}\rangle_{e}.$).
It is expected that in the limit of weak backaction (small $\phi$), the nuclear spin is rotated in $\sigma_X-\sigma_Y$ plane, resulting in oscillations of these probabilities with a frequency of $2\delta.$
The accumulative effect of the backaction becomes significant for $g^{2}\tau_{m} t>1,$ therefore clear oscillations of the nuclear spin can be seen only for $\omega>g^{2}\tau_{m}.$ (a) Oscillations of the probability for $g=\delta=0.05\tau^{-1}.$ These probabilities satisfy: $p_{\pm}\approx0.5\pm0.5\sin\left(2g\tau_{m}\right)\cos\left(2\omega t\right).$
Whereas in (b): The probability for the same $g,$ and $\delta=0.01g=5\cdot 10^{-4} \tau^{-1}.$ In that regime almost no oscillations can be observed.  }

\label{NMR_behavior}
\end {center}
\end{figure}

 Let us first focus on the case of a {\it{polarized}} nuclear spin, and examine the behavior of
the FI. The information is extracted from the weak measurement, however
these weak measurements affect the rotation of the nuclear spin as they induce back
action. Note that as long as $g^{2}\tau_{m} T\ll1,$ the effect
of the backaction can be neglected, and the oscillations persist irrespective
of the weak measurements. Therefore in the limit of $g^{2}\tau_{m} T\ll1,$
we can neglect the effect of the back-action, and thus the probability
of collapsing into into $|\uparrow_{y}\rangle_{e}/|\downarrow_{y}\rangle_{e}$
reads: 
\begin{eqnarray*}
\begin{split}
p_{\pm} & \approx \cos^{2}\left(g\tau_{m}\pm\frac{\pi}{4}\right)\cos^{2}\left(\delta t\right)+\cos^{2}\left(g\tau_{m}\mp\frac{\pi}{4}\right)\sin^{2}\left(\delta t\right)\\
 & =0.5\pm0.5\sin\left(2g\tau_{m}\right)\cos\left(2\delta t\right),
 \end{split}
\end{eqnarray*}
 as is illustrated in fig. \ref{NMR_behavior}. 
 
Given these probabilities, the FI reads:
\begin{equation}
I\approx\underset{t}{\sum}4t^{2}\sin\left(2g\tau_{m}\right)^{2}\sin\left(2\delta t\right)^{2}\approx16\phi^{2}\underset{t}{\sum}t^{2}\sin\left(2\delta t\right)^{2}.
\end{equation}
This expression of the FI is valid only for $T$ satisfying $g^{2}\tau_{m}T \ll 1,$ which is the weak measurement regime.

\begin{figure}
\begin {center}
\subfigure[]{\includegraphics[width=6.5cm]{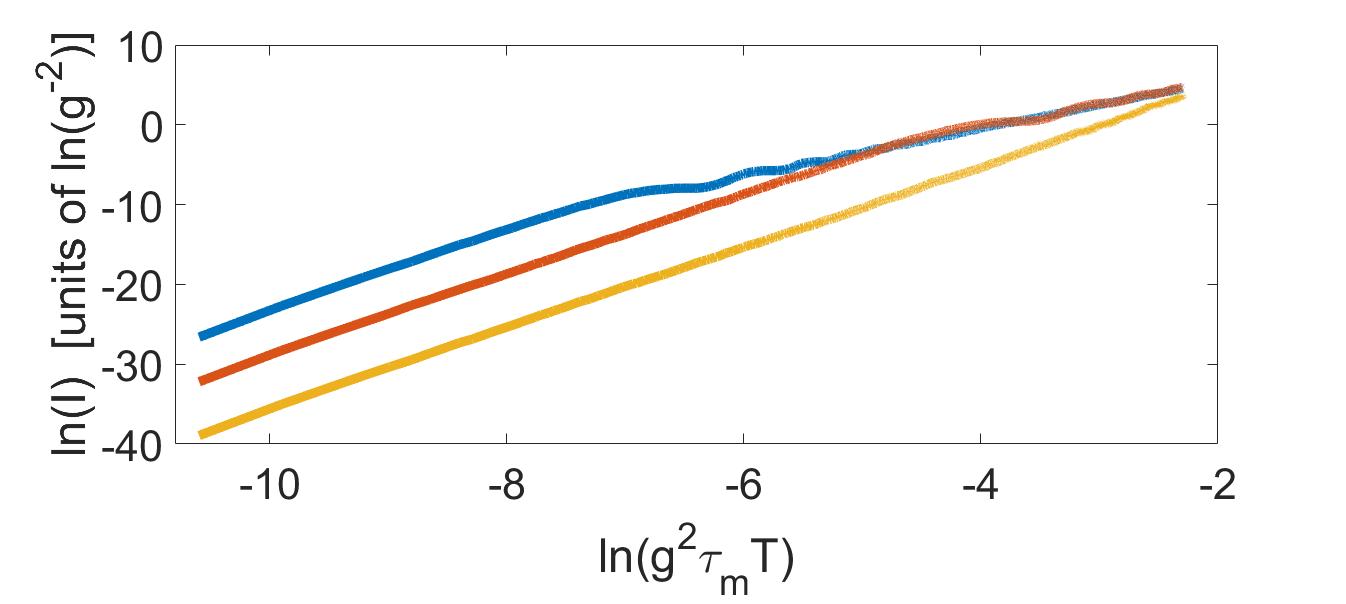}}
\subfigure[]{\includegraphics[width=6.5cm]{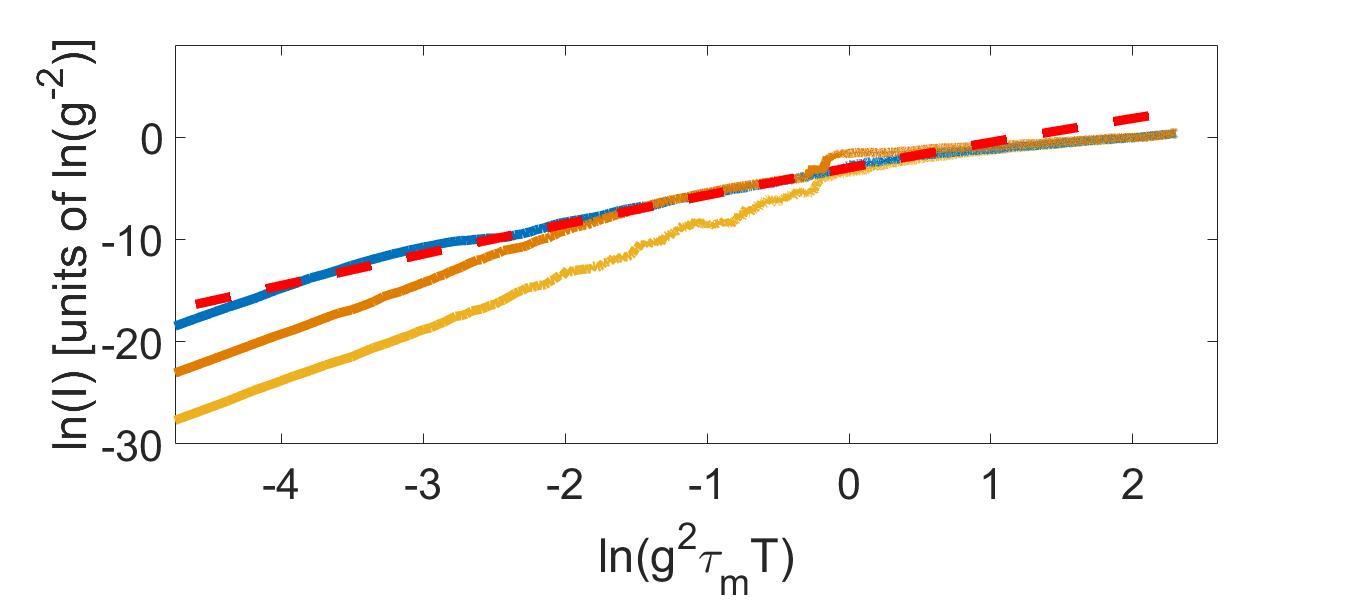}}
\caption{The FI achieved with NMR method for polarized nuclear spin, as a function of time, for different detunings. (a) $g=5\cdot10^{-3}\left[\tau_{m}^{-1}\right],$ and $\delta=5\:\text{(blue,upper curve)},0.1\:\text{(red,middle curve)},0.01\:\text{(yellow,lower curve)}$ (in units of $g$).
It can be seen that for short times ($\delta T \ll 1$), $I\propto g^{2}\tau_{m}\delta^{2}T^{5}.$ Therefore, in that period, the smaller $\delta,$ the smaller the FI, as seen in the figure. 
For $\delta^{-1}<T<\left(g^{2}\tau_{m}\right)^{-1},$ $I\propto g^{2}\tau_{m} T^{3},$ so that in this period the FI does not depend on $\delta.$ 
 (b) For $T>\left(g^{2}\tau_{m}\right)^{-1},$ the behavior of the FI is changed, and the scaling converges to $T,$ due to the accumulated effect of the backaction. 
 The dashed red curve corresponds to the FI achieved with the method described in the main text (optimized over the number of measurements). For $\delta^{-1}<T<\left(g^{2}\tau_{m}\right)^{-1},$ both methods achieve a similar performance, but for $T>\left(g^{2}\tau_{m}\right)^{-1},\, T<{\delta^{-1}}$ the method described in the main text outperforms the NMR method. }
\label{FI_comparison_NMR}
\end {center}
\end{figure}

Note that this is exactly the same FI achieved when sensing a classical signal (which is not surprising as we neglected backaction), where $g$ is analogous to the amplitude of the classical signal.
One can see that for very short times, $\delta T\ll1$, we have $I\approx\frac{64}{5}g^{2}\tau_{m}\delta^{2}T^{5}.$
Namely the FI drops as $\delta$ gets smaller, and in particular when $\delta\rightarrow0$ the FI vanishes.
This is just due to the fact that the oscillations start in the antinodes, so that the derivative according to $\delta$ vanishes. It should be noted.  
This implies that working with a small $\delta$ (compared to $1/T$) leads to a poor FI and should be thus avoided. In fact, this can be also solved by applying a pulse on the nuclear spin at the beginning, which will rotate its polarization to $\sigma_{y}$ basis.
Given that its initial state is in $\sigma_{y}$ basis and we measure weakly in $\sigma_{x}$ basis, the oscillations now start in the nodes and the FI does not vanish. 
For longer times, $\delta T >1$ (but still in the weak measurement regime) we get: $I\approx\frac{8}{3}g^{2}\tau_{m} T^{3},$ Recall that the FI of our method behaves in a similar manner at this regime, as it also goes as $g^{2}\tau_{m}T^{3}.$
In fact, in this regime the NMR method seems to outperform our method by a factor (which is of order 1).   
This behavior of the FI is shown in figure \ref{FI_comparison_NMR}.

For longer times, $g^{2}\tau_{m} T>1,$ the strong measurement regime, the behavior is changed, 
and the scaling of the FI converges to $T$ (analytical form is unknown). In this regime, the FI hardly dependes on $\delta$. The crucial point is that in the strong measurement regime the FI of the NMR does not get close to $4 T^{2},$ as the backaction ruins the accumulation of the phase.
Thus in this regime our method significantly outperforms the NMR method. This analysis is illustrated in fig. \ref{FI_comparison_NMR}.\\

\begin{figure}
\begin {center}
\subfigure[]{\includegraphics[width=6.9cm]{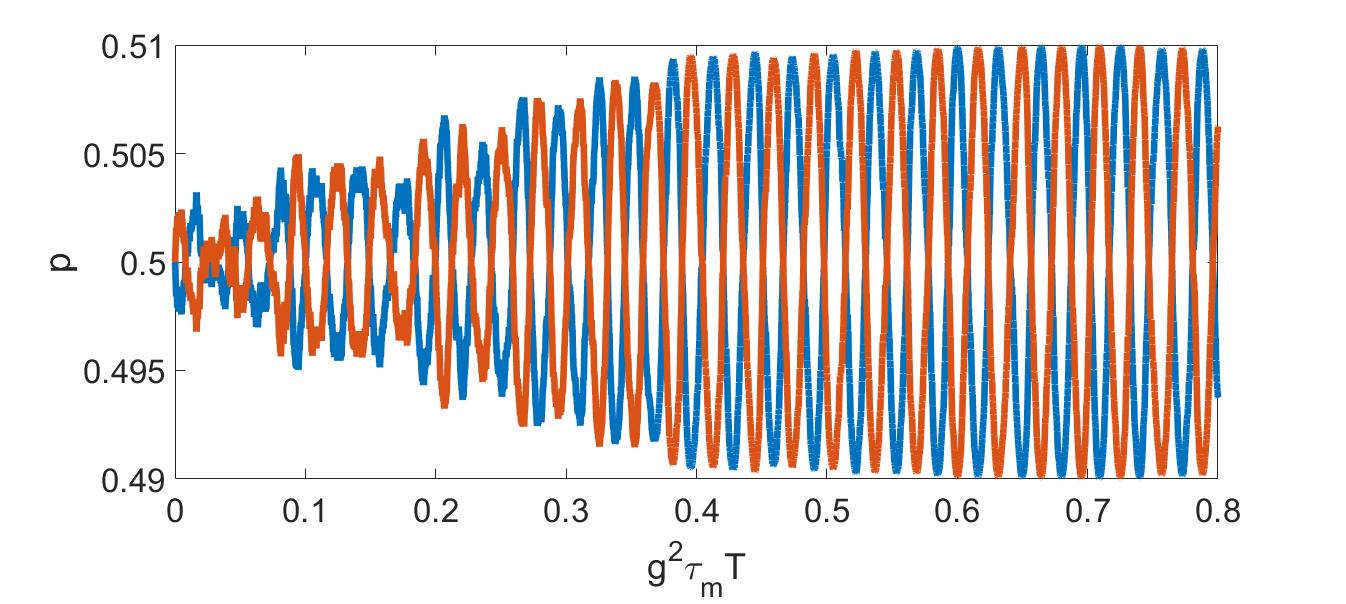}}
\subfigure[]{\includegraphics[width=6.5cm]{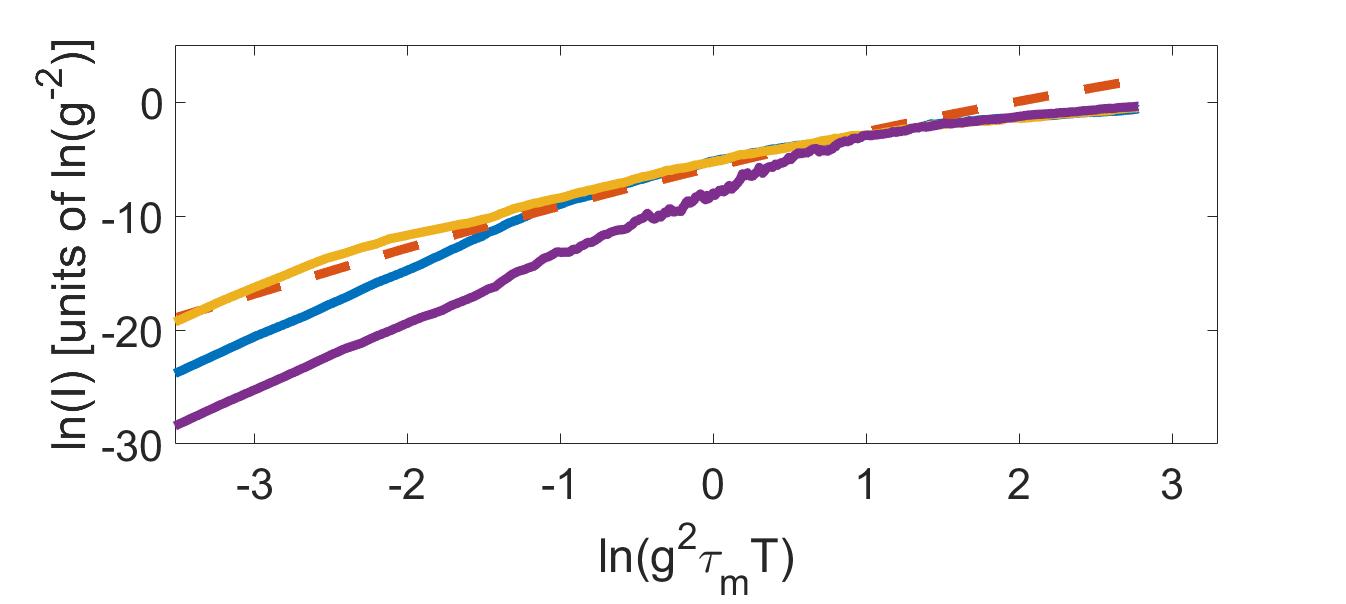}}       
\caption{NMR method for unpolarized nuclear spin. (a) Dynamics of the nuclear spin as a function of time. The different curves correspond to $p_{+},\, p_{-}.$ For $g^{2}\tau_{m}t \ll 1$ the envelope of the oscillations grows due to the increasing polarization. For longer times the nuclear spin becomes fully polarized. The dynamics then coincide with that of the polarized case, and the FI is also the same as in the polarized case.
(b) FI for different detunings compared to the FI of our method. The solid yellow (upper), blue (middle) and purple (bottom) curves correspond to NMR method with $\delta=1,0.1,0.01 \, \left[g\right]$ respectively.
The dashed orange curve correspond to our method.  }

\label{FI_comparison_NMR_unpolarized}
\end {center}
\end{figure}

Regarding {\it{unpolarized}} nuclear spin, the dynamics is more involved: The polarization grows due to the measurements along with the oscillations, as can be seen in figure \ref{FI_comparison_NMR_unpolarized}.
Therefore for $T<\left(g^{2}\tau_{m}\right)^{-1},$ the behavior of the FI is different from that of the polarized (and of course the FI is much smaller), for $T>\left(g^{2}\tau_{m}\right)^{-1},$ the nuclear spin is polarized ($p_\text{pol}=1$), and the FI coincides with that of the polarized (as can be seen in fig. \ref{nmr_main_text}).
As with the polarized scenario, our method outperforms the NMR in case of a long time (compared to measurement strength), namely $g^{2}\tau_{m}T\gg1.$ In the weak measurement regime, these two methods coincide.  

The FI given by the NMR method in the strong measurement regime is damaged due to the backaction induced by the measurements. The required number of measurements to polarize and measure the nuclear spin strongly goes as $\left(g\tau_{m}\right)^{-2},$ additional measurements just ruin the oscillations and do not provide further information.
Therefore in this regime, gaps between consecutive measurements (in the  NMR method) will definitely increase the FI. This will reduce the total number of measurements and the nuclear spin will rotate freely during these gaps.
However this will still not be optimal, in order to make the phase accumulation of the nuclear spin more efficient the best is to stick all these gaps together, just like it is done in the method we propose.           

\section{Derivation of the probability function}
The full result of the probability function can also be derived without assuming weak coupling. 
We define the unitaries  that act on the nucleus space conditioned on the NV state of $\ket{\uparrow_z}$ and $\ket{\downarrow_z}$ at $t=0$  by $U_{\pm}$ and  at $t=T$  by $V_{\pm}.$
\begin{align}
U_{\pm} = \Pi_{k=1}^N U_{\pm,k}\, , V_{\pm} = \Pi_{k=1}^N V_{\pm,k},
\end{align}
where
\begin{align}
U_{\pm,k} & = \cos \left( N_k \tau \right) \, \mathbb{I} - i \frac{\sin(N_k\tau)}{N_k}\left(  \mp g_k I_x + \delta_k  I_z  \right),
\notag \\
V_{\pm,k} & = \cos \left( N_k \tau \right)  \, \mathbb{I} - i \frac{\sin(N_k\tau)}{N_k}\left(  \mp g_k (I_x \cos (\omega_k T) + I_y \sin (\omega_k T)) + \delta_k I_z \right),
\end{align}
are the unitaries of the individual nuclei
and
\begin{align}
N_k^2 = \delta_k^2 + g_k^2.
\end{align}
The postselected state after getting two $\uparrow_x$ results is:
\begin{equation}
\frac{1}{4} \left(V_+ +V_-  \right)  \left(U_+ + U_-  \right) \vert \psi_N  \rangle.
\end{equation}
Given that the initial state of the nuclei is the identity matrix the exact expression of the probability function that we get is:
\begin{eqnarray}
\begin{split}
&p_{\uparrow_x,\uparrow_x}  = \frac{ 1}{ 2^N} \frac{1 }{16} 
\Big[2^N  4  + \Pi_{k=1}^N \mathrm{Tr} \left[ V_{+,k} V_{-,k}^{\dag} U_{+,k}^{\dag} U_{-,k} \right] +\\
 &\Pi_{k=1}^N \mathrm{Tr} \left[ V_{-,k} V_{+,k}^{\dag} U_{-,k}^{\dag} U_{+,k} \right]
-  \Pi_{k=1}^N \mathrm{Tr} \left[ V_{+,k} V_{-,k}^{\dag} U_{-,k}^{\dag} U_{+,k} \right]
\notag \nonumber\\
&  \Pi_{k=1}^N \mathrm{Tr} \left[ V_{-,k} V_{+,k}^{\dag} U_{+,k}^{\dag} U_{-,k} \right]\Big].
\end{split}
\end{eqnarray}
Where the traces are equal to: 

{\small 
\begin{eqnarray}
\begin{split}
& \mathrm{Tr} \left[ V_{+,k} V_{-,k}^{\dag} U_{+,k}^{\dag} U_{-,k} \right] = \mathrm{Tr} \left[ V_{-,k} V_{+,k}^{\dag} U_{-,k}^{\dag} U_{+,k} \right] =\nonumber\\
& \frac{2 \delta_k ^4+g^4}{\left(\delta_k ^2+g_k^2\right)^2}+\\
 &\frac{g_k^2 \left(8 \delta_k ^2 \cos ^2\left(\frac{\omega_k  T}{2}\right) \cos \left(2 \tau
   N_k\right)+\cos \left(4 \tau N_k\right) \left(g_k^2-\left(2 \delta_k
   ^2+g_k^2\right) \cos (\omega_k  T)\right)\right)}{\left(\delta_k ^2+g_k^2\right)^2}  \nonumber\\
 & - \frac{g_k^2 \left(16 \delta_k  N_k \sin (\omega_k  T) \sin ^3\left(\tau
  N_k\right) \cos \left(\tau N_k \right)+g_k^2 \cos (\omega_k  T)-2 \delta_k ^2 \cos
   (\omega_k  T)\right)}{\left(\delta_k ^2+g_k^2\right)^2}\nonumber\\
\end{split}
\end{eqnarray} }

{\small 
\begin{eqnarray}
&& \mathrm{Tr} \left[ V_{+,k} V_{-,k}^{\dag} U_{-,k}^{\dag} U_{+,k} \right] = \mathrm{Tr} \left[ V_{-,k} V_{+,k}^{\dag} U_{+,k}^{\dag} U_{-,k} \right] =\nonumber\\
&&\Big(\frac{-4 \delta_k ^2 g_k^2 (\cos (\omega_k  T)-1) \cos \left(2 \tau \sqrt{\delta_k ^2+g_k^2}\right)}{\left(\delta_k ^2+g_k^2\right)^2}\nonumber \\ 
&&+\frac{g_k^2 \cos (\omega_k  T)
   \left(2 \delta_k ^2+\left(2 \delta_k ^2+g_k^2\right) \cos \left(4 \tau N_k\right)-g_k^2\right)} {\left(\delta_k ^2+g_k^2\right)^2}\times\nonumber\\
   &&+\frac{ 2 \delta_k  \left(\delta_k ^3+8 g_k^2 N_k \sin (\omega_k  T) \sin ^3\left(\tau N_k\right) \cos \left(\tau N_k\right)\right)+g_k^4 \left(\cos \left(4 \tau N_k\right)+1\right)}{\left(\delta_k ^2+g_k^2\right)^2}\Big)\times \nonumber\\
  &&\frac{g_k^4 \left(\cos \left(4 \tau N_k\right)+1\right)}{\left(\delta_k ^2+g_k^2\right)^2}\nonumber\\
  \end{eqnarray} }

\baselineskip=12pt
\bibliography{Bib_Alex}
\bibliographystyle{apsrev4-1}

\end{document}